\documentclass[a4paper,11pt]{article}

\usepackage{amsmath,amsfonts,amssymb,caption,graphicx,natbib}
\input{epsf}

\def\trev{\text{rev}}

\newcommand{\weakly}{\mbox{$ \;\stackrel{\cal D}{\longrightarrow}\; $}}

\newcommand{\tod}{\mbox{$ \;\stackrel{\cal D}{\to}\; $}}

\newcommand{\RL}{{\mathbb R}}

\newcommand{\IND}{{\mathbb I}}

\newcommand{\K}{{\rm K}}
 
\newcommand{\VAR}{\mbox{\rm Var}}

\newcommand{\COV}{\mbox{\rm Cov}}

\newcommand{\iid}{\mbox{i.i.d.}\!}

\def\be{\begin{eqnarray}}
\def\ee{\end{eqnarray}}
\def\ben{\begin{eqnarray*}}
\def\een{\end{eqnarray*}}

\def\flabel#1{\label{f:#1}}
\def\elabel#1{\label{e:#1}}

\def\sq{$\Box$}

\def\qed{\ifmmode\sq\else{\unskip\nobreak\hfil
\penalty50\hskip1em\null\nobreak\hfil\sq
\parfillskip=0pt\finalhyphendemerits=0\endgraf}\fi\par\medbreak}

\newsavebox{\junk}
\savebox{\junk}[1.6mm]{\hbox{$|\!|\!|$}}

\def\det{{\mathop{\rm det}}}

\def\state{{\sf X}}

\def\til={{\widetilde =}}

\def\clB{{\cal B}}

 \def\eq#1/{(\ref{#1})}

\def\Section#1{Section~\ref{s:#1}}

\def\eq#1/{(\ref{e:#1})}

\newcommand{\beqn}[1]{\notes{#1}%
\begin{eqnarray} \elabel{#1}}

\newcommand{\eeqn}{\end{eqnarray} }

\newcommand{\beq}[1]{\notes{#1}%
\begin{equation}\elabel{#1}}

\newcommand{\eeq}{\end{equation}} 

\def\bdes{\begin{description}}
\def\edes{\end{description}}

\def\notes#1{}

\begin{document}

\title{\vspace{-1.5cm}%
Control Variates for Reversible MCMC Samplers}

\author
{
    Petros Dellaportas
    \thanks{Department of Statistics,
        Athens University of Economics and Business
	(Kontrikton campus),
        Patission 76, Athens 10434, Greece.
                Email: {\tt petros@aueb.gr}.
        }
\and
        Ioannis Kontoyiannis
    \thanks{Department of Informatics,
        Athens University of Economics and Business
	(Kontrikton campus),
        Patission 76, Athens 10434, Greece.
                Email: {\tt yiannis@aueb.gr}.
        }
    \thanks{I.\ Kontoyiannis was supported in part 
	by a Marie Curie International
	Outgoing Fellowship, PIOF-GA-2009-235837.
	}
}

\date{\today}

\maketitle

\begin{abstract}
A general methodology is introduced for the construction and
effective application of control variates to estimation problems
involving data from reversible MCMC samplers. We propose the use 
of a specific class of functions as control variates, and we 
introduce a new, consistent estimator for the values of the 
coefficients of the optimal linear combination of these functions.
The form and proposed construction of the control variates 
is derived from our 
solution of the Poisson equation associated with a specific 
MCMC scenario. The new estimator, which can be applied to 
the same MCMC sample, is derived from a novel, finite-dimensional,
explicit representation for the optimal coefficients.
The resulting variance-reduction methodology is primarily 
applicable when the simulated data are generated by a
conjugate random-scan Gibbs sampler. MCMC examples 
of Bayesian inference problems demonstrate that the corresponding
reduction in the estimation variance is significant,
and that in some cases it can be quite dramatic. 
Extensions of this methodology in several directions are 
given, including certain families of Metropolis-Hastings 
samplers and hybrid Metropolis-within-Gibbs algorithms. 
Corresponding simulation examples are presented illustrating 
the utility of the proposed methods. All methodological and 
asymptotic arguments are rigorously justified under easily 
verifiable and essentially minimal conditions.
\end{abstract}



\thispagestyle{empty}
\setcounter{page}{0}

\newpage

\section{Introduction}
\label{s:intro}

Markov chain Monte Carlo (MCMC) methods provide the facility to
draw, in an asymptotic sense, a sequence of dependent samples
from a very wide class of probability measures in any dimension. 
This facility, together with the tremendous 
increase of computer power in recent years, makes MCMC 
perhaps the main reason for the widespread use of Bayesian 
statistical modeling and inference across the spectrum 
of quantitative scientific disciplines.

This work provides a methodological foundation for
the construction and use of control variates in conjunction
with reversible MCMC samplers. Although popular in the 
standard Monte Carlo setting, control variates 
have received much less attention in the MCMC literature.
The proposed methodology will be shown, both via 
theoretical results and simulation examples, 
to reduce the variance of the resulting 
estimators significantly, and sometimes 
quite dramatically.

In the simplest Monte Carlo setting, when
the goal is to compute the expected value 
of some function $F$ evaluated on independent 
and identically distributed
(i.i.d.) samples $X_1,X_2,\ldots$, the variance 
of the standard ergodic averages of the $F(X_i)$
can be reduced by exploiting available zero-mean 
statistics. If there are one or more functions 
$U_1,U_2,\ldots,U_k$ --
the {\em control variates} -- for which it
is known that the expected value of each $U_j(X_1)$
is equal to zero, then subtracting any linear 
combination, $\theta_1U_1(X_i)+
\theta_2U_2(X_i)+\cdots+
\theta_kU_k(X_i),$
from the $F(X_i)$ does not change 
the asymptotic mean of the corresponding
ergodic averages. 
Moreover,
if the best constant coefficients $\{\theta^*_j\}$ 
are used, then the variance
of the estimates is no larger than before and
often it is much smaller. The standard practice
in this setting is to estimate the
optimal $\{\theta_j^*\}$ based on the
same sequence of samples;
see, e.g., \citet{liu:book}, 
\citet{robert:book}, or
\citet{givens:book}.
Because of the demonstrated effectiveness
of this technique, in many important areas
of application, e.g., in computational
finance where Monte Carlo methods are a basic
tool for the approximate computation 
of expectations, see \citet{glasserman:book},
a major research effort has been devoted to
the construction of effective control variates
in specific applied problems.

The main difficulty in 
extending the above methodology to
estimators based on MCMC samples is probably
due to the intrinsic complexities presented 
by the Markovian structure. 
On one hand it is hard to find nontrivial, 
useful functions
with known expectation with respect to the stationary
distribution of the chain\footnote{For example, 
	\citet{mr1} comment
	that ``control variates have been advertised early in the MCMC
	literature (see, e.g., \citet{greenhan}), but they 
	are difficult to work 
	with because the models are
	always different and their complexity is such that it is
	extremely challenging to
	derive a function with known expectation.''}, 
and, even in cases
where such functions are available, there has been
no effective way to obtain consistent
estimates of the corresponding optimal coefficients 
$\{\theta_j^*\}$. An important underlying reason
for both of these difficulties is the basic fact 
that the MCMC variance of ergodic averages is
intrinsically an infinite-dimensional object:
It cannot be expressed in closed form as a function 
of the transition kernel and the stationary distribution
of the chain.

An early reference of variance reduction for Markov chain samplers
is \citet{greenhan}, who exploit an idea of \citet{barfri89}
and construct antithetic variables that may achieve variance
reduction in simple settings but do not appear to
be widely applicable.
\citet{adra} focus on finite-state space chains,
they observe that optimum variance reduction can be
achieved via the solution of the associated 
{\em Poisson equation} (see equation~(\ref{eq:PoissonIntro}) 
below and \Section{general} for details), 
and they propose numerical algorithms 
for its solution.
Rao-Blackwellisation has been suggested by \citet{gelfand+smith} and
by \citet{robert:book} as a way to reduce the variance of MCMC
estimators. Also, \citet{Philrob} investigated the use of Riemann 
sums as a variance reduction tool in MCMC algorithms. An interesting as
well as natural control variate that has been used, mainly as
a convergence diagnostic, by \citet{FanBrGe}, is the score statistic.
Although \citet{Philrob} mention that it can be used as a control
variate, its practical utility has not been investigated.
\citet{Atchade_perron} restrict attention to independent Metropolis
samplers and provide an explicit formula 
for the construction of control variates.
\citet{Mira_tecrep} note that best control variate
is intimately connected with the solution of the
Poisson equation and they
attempt to solve it numerically. \citet{Hammer08} construct control
variates for general Metropolis-Hastings samplers by expanding the
state space. 

In most of the works cited above,
the method used for the estimation of the optimal coefficients 
$\{\theta_j^*\}$ is either based on the same formula as the one
obtained for control variates in i.i.d.\ Monte Carlo sampling,
or on the method of {\em batch-means},
but such estimators are strictly suboptimal and generally
ineffective; 
see Section~\ref{s:Ifurther} for a summary
and Section~\ref{s:batch} for details.

For our purposes, a more relevant line of work is
that initiated by \citet{henderson:phd},
who observed that, for {\em any} real-valued 
function $G$ defined on the state space of 
a Markov chain $\{X_n\}$, the function
$U(x):=G(x)-E[G(X_{n+1})|X_n=x]$
has zero mean with respect to the stationary 
distribution of the chain. 
\citet{henderson:phd}, like some of the other
authors mentioned above, also notes that the
best choice for the function $G$ would
be the solution of the associated 
Poisson equation
and proceeds to compute
approximations of this solution for specific
Markov chains, with particular emphasis
on models arising in stochastic network
theory.

The gist of our approach is to 
adapt Henderson's
idea and use the resulting control variates
in conjunction with a new,
efficiently implementable and provably optimal
estimator for the coefficients $\{\theta_j^*\}$.
The ability to estimate
the $\{\theta_j^*\}$ effectively makes these
control variates practically relevant in
the statistical MCMC context, and avoids 
the need to
compute analytical approximations to the
solution of the underlying Poisson equation.

\subsection{Outline of the proposed basic methodology}
\label{s:outline}

Section~\ref{s:general} introduces the general
setting within which all the subsequent results
are developed. A sample of size $n$
from an ergodic Markov chain $\{X_n\}$
is used to estimate the mean $E_\pi(F)=\int F\,d\pi$
of a function $F$, under the unique invariant
measure $\pi$ of the chain. The associated
{\em Poisson equation} is introduced, and 
it is shown that its solution can be used 
to quantify, in an essential way, the rate at 
which the chain converges to equilibrium.

In Section~\ref{s:cv2}
we examine the variance of the standard ergodic
averages,
\be
\mu_n(F):=\frac{1}{n}\sum_{i=0}^{n-1}F(X_i),
\label{eq:Iergodic}
\ee 
and we
compare it with the variance of the
modified estimators,
\be
\frac{1}{n}\sum_{i=1}^n\big[F(X_i)-\theta_1U_1(X_i)
-\theta_2U_2(X_i)
-\cdots-\theta_kU_k(X_i)\big].
\label{eq:intro-mod}
\ee
Here and throughout the subsequent discussion,
the control variates $U_1,U_2,\cdots,U_k$,
are constructed as above via,
$U_j(x):=G_j(x)-PG_j(x)$, where 
$PG(x)$ denotes the one-step expectation
$E[G(X_{n+1})|X_n=x]$,
for particular choices of the functions
$G_j$, $j=1,2,\ldots,k$.

The two central methodological issues 
addressed in this work are: (i)~The 
problem of estimating the 
optimal coefficient vector $\{\theta^*_j\}$ 
that minimizes the variance of the modified
estimates~(\ref{eq:intro-mod});
and (ii)~The choice of the functions $\{G_j\}$, 
so that the corresponding functions $\{U_j\}$
will be effective as control variates 
in specific MCMC scenarios that arise
from common families
of Bayesian inference problems.

For the first issue, 
in Section~\ref{s:thetastar}, 
we derive new representations 
for the optimal coefficient vector 
$\{\theta_j^*\}$, under the assumption
that the chain $\{X_n\}$ is {\em reversible};
see Proposition~2.
These representations lead to 
our {\em first main result}, namely,
a new estimator for 
$\{\theta_j^*\}$; see equations~(\ref{eq:theta-n-K}) 
and~(\ref{eq:theta-optimal-est}).
This estimator 
is based on the same MCMC output and it can 
be used after the sample has been obtained, 
making its computation independent 
of the MCMC algorithm used.

The second problem, that of selecting 
an effective collection of functions
$\{G_j\}$ for the construction of the
control variates $\{U_j\}$, is more complex
and it is dealt with in stages.
First, in Section~\ref{s:cv2}
we recall that there is always
a single choice of a function $G$
that actually makes the estimation 
variance equal to zero:
If $G$ satisfies,
\be
U:=G-PG=F-E_\pi(F),
\label{eq:PoissonIntro}
\ee
then with this control variate and with $\theta=1$
the modified estimates
in (\ref{eq:intro-mod})
are {\em equal} to the required
expectation $E_\pi(F)$ for 
{\em all} $n$.
A function $G$ satisfying (\ref{eq:PoissonIntro})
is often called a solution to the
{\em Poisson equation for $F$}
[or {\em Green's function}].
But solving the Poisson equation
even for simple functions $F$ 
is a highly nontrivial task, and for 
chains arising in typical applications
it is, for all practical purposes,
impossible;
see, e.g., the relevant
comments in \citet{henderson:phd} 
and \citet{meyn:book}.
Therefore, as a first {\em rule of thumb},
we propose that a class
of functions $\{G_j\}$ be chosen
such that the solution to the
Poisson equation (\ref{eq:PoissonIntro})
can be accurately approximated by a linear 
combination $\sum_j\theta_jG_j$ of
the $\{G_j\}$. For this reason 
we call the $\{G_j\}$
{\em basis functions}.

Clearly there are many possible choices
for the basis functions
$\{G_j\}$, and the effectiveness of the
resulting control variates depends
heavily on their particular choice. 
In order to obtain more specific and
immediately-applicable guidelines for the 
selection of the $\{G_j\}$, in
Section~\ref{s:basic} we note
that there are two basic requirements
for the immediate applicability of the methodology
described so far; the underlying chain
needs to be reversible in order for the
estimates of the coefficient vector
$\{\theta_j^*\}$ introduced in 
Section~\ref{s:thetastar} to be consistent,
and also,
the one-step expectations
$PG_j(x):=E[G_j(X_{n+1})|X_n=x]$ that are necessary
for the construction of the control variates
$U_j$ need to be explicitly computable.

Since the most commonly used class of MCMC 
algorithms satisfying both of these
requirements is that of conjugate random-scan
Gibbs samplers,\footnote{Following standard
	parlance, we call a Gibbs sampler `conjugate'
	if the full conditionals of the target distribution
	are all known and of standard form. This, of course,
	is unrelated to the notion of a conjugate prior
	structure in the underlying Bayesian formulation.}
and since the most accurate
general approximation of the target distribution
$\pi$ arising in Bayesian inference problems
is a general multivariate Gaussian,
we examine this MCMC problem in detail
and obtain our {\em second main result}.
Suppose that we wish to 
estimate the mean of one of the co-ordinates
of a $k$-dimensional Gaussian distribution $\pi$, 
based on samples 
$X_i=(X_i^{(1)},X_i^{(2)},\ldots,X_i^{(k)})^t$
generated by the
random-scan Gibbs algorithm. In Theorem~1
we show that the solution of the associated
Poisson equation can always be expressed as 
a linear combination of the $k$ co-ordinate
functions $G_j(x):=x^{(j)}$, 
$x=(x^{(1)},x^{(2)},\ldots,x^{(k)})^t\in\RL^k$.
This is perhaps the single most interesting 
case of a Markov chain naturally arising in 
applications for which an explicit solution
to the Poisson equation has ever been obtained.

The above results naturally lead to the following
proposal, stated in detail in Section~\ref{s:basic}.
It is the core methodological contribution of this
work.

\begin{quote}
{\em The basic methodology.}
Suppose $\pi$ is a multivariate posterior distribution 
for which MCMC samples are obtained by a reversible Markov 
chain $\{X_n\}$. In order to estimate the posterior mean 
$\mu^{(i)}$ of the $i$th co-ordinate $x^{(i)}$,
let $F(x)=x^{(i)}$, define
basis functions $G_j$ as the 
co-ordinate functions $G_j(x)=x^{(j)}$ for all components $j$ 
for which $PG_j(x)=E[X_{n+1}^{(j)}\,|\,X_n=x]$ is 
explicitly computable, and form the control variates 
$U_j=G_j-PG_j$. 

Then estimate the optimal coefficient 
vector $\theta^*=\{\theta_j^*\}$ by the 
estimator $\hat{\theta}=\hat{\theta}_{n,\K}$ 
given in~(\ref{eq:theta-n-K}), and estimate the 
posterior mean of 
interest $\mu^{(i)}$ by the modified 
estimators given in (\ref{eq:multi-est2}):
\be
\mu_{n,\K}(F):=
\frac{1}{n}\sum_{i=1}^n\big[F(X_i)-
\hat{\theta}_1U_1(X_i)
-\hat{\theta}_2U_2(X_i)
-\cdots
-\hat{\theta}_kU_k(X_i)\big].
\label{eq:Imod}
\ee
\end{quote}

Section~\ref{s:simple} contains three MCMC examples 
using this methodology. Example~1 is a brief illustration 
of the result of Theorem~1 in the case of a bivariate
Gaussian. As expected, the modified estimators
(\ref{eq:Imod}) are seen to be much more effective 
than the standard ergodic averages~(\ref{eq:Iergodic}),
in that their variance is smaller by a factor ranging 
approximately between 4 and 1000, depending on the sample 
size.
Example~2 contains an analysis of a realistic 
Bayesian inference problem via MCMC, 
for a 66-parameter hierarchical normal 
linear model. There, we consider all 66 problems
of estimating the posterior means of all 66
parameters, and we find that in most cases
the reduction in variance resulting from the use 
of control variates as above is typically
by a factor ranging between 5 and 30.
The third example illustrates the use of the
basic methodology in the case of 
Metropolis-within-Gibbs sampling
from a heavy-tailed posterior. Even
though the posterior distribution 
is highly non-Gaussian, and also
the one-step expectation $PG_j$
can be computed for only one of the two
model parameters, we still find that
the variance is reduced by a factor ranging
approximately between 7 and 10.

\subsection{Extensions and applications}

{\em Domain of applicability. }
The present development not only generalizes 
the classical method of control variates to the MCMC 
setting, but it also offers an important advantage.
In the case of independent sampling, 
the control variates for each specific application 
need to be identified from scratch, often
in an {\em ad hoc} fashion. In fact, for most
Monte Carlo estimation problems there are
{\em no} known functions that can be used
as effective control variates.
In contrast, the basic methodology described above 
provides a way of constructing a family 
of control variates that
are immediately applicable to a wide range 
of MCMC problems,
as long as the sampling algorithm
produces a reversible chain for which the one-step
expectations $PG(x):=E[G(X_{n+1})|X_n=x]$ can be
explicitly computed for some simple, linear 
functions $G$.
MCMC algorithms with these properties form
a large collection of samplers commonly used 
in Bayesian inference, including, among others:
All conjugate random-scan Gibbs samplers
(the main MCMC class considered in this work),
certain versions of 
hybrid Metropolis-within-Gibbs algorithms, 
and certain types of Metropolis-Hastings
samplers on discrete states spaces.

In Section~\ref{s:generalMeth} we discuss extensions
of the basic methodology along three directions
that, in some cases, go beyond the above class
of samplers. 

\medskip

\noindent
{\em General classes of basis functions $\{G_j\}$. }
Section~\ref{s:generalMeth} begins with
the observation that, although it is often effective 
to take the basis functions $\{G_j\}$ 
to be the co-ordinate functions of the chain,
in certain applications it is natural to consider 
different families of $\{G_j\}$.
Sometimes it is the structure of the MCMC sampler 
itself that dictates a more suitable choice,
while sometimes,
in estimation problems involving more complex 
models -- 
particularly in cases where it is expected that the 
posterior distribution is highly non-Gaussian -- 
it may be possible to come up with more effective 
basis functions $\{G_j\}$ by considering the form
of the associated Poisson equation, as indicated by
the first rule of thumb stated earlier.
Finally, when the statistic of interest is not the 
posterior mean of one of the model parameters, 
there is little reason to expect that the co-ordinate
functions $\{G_j\}$ would still lead to effective
control variates; instead, we argue that it is more 
appropriate to choose functions $G_j$ leading to 
control variates $U_j$ that are highly correlated
with the statistic of interest.

Particular instances of these three directions
are illustrated via the three MCMC examples presented 
in Section~\ref{s:generalMeth} and summarized below.

\medskip

\noindent
{\em `Non-conjugate' samplers. }
In Example~4, a non-conjugate Gibbs sampler
is used for the analysis of the posterior of 
a 7-parameter log-linear model. There, 
the conditional expectations 
$PG_j$ of the co-ordinate functions
$G_j$ are not available in closed form, so that 
the basic methodology cannot be applied. 
But the nature of the sampling algorithm 
provides a natural and in a sense obvious
alternative choice for the basis functions,
for which the computation of the 
functions $PG_j$ needed for the construction
of the associated control variates $U_j=G_j-PG_j$ 
is straightforward. With this choice of control
variates, using the modified estimator
in (\ref{eq:Imod}) to estimate the posterior 
means of the seven model parameters, we find 
that the variance is approximately between 3.5 
and 180 times smaller than that of the standard 
ergodic averages~(\ref{eq:Iergodic}).

\medskip

\noindent
{\em Complex models. }
Example~5 is based on a two-component Gaussian
mixtures model, endowed with the vague prior 
structure of \citet{rich}. To estimate the posterior
means of the two parameters corresponding
to the Gaussian locations, we use a 
random-scan Gibbs sampler (in blocks).
Here, although the prior specification 
is conditioned on the two means being ordered,
it is possible to generate samples 
without the ordering restriction, and then
to order the samples at the post-processing
stage. The resulting Markov chain has an 
apparently quite complex structure, yet we
show that the basic methodology can easily
be applied in this case. What makes this
example interesting for our present purposes
is that, on one hand, the control variates 
based on the co-ordinate basis functions $\{G_j\}$
make little (if any) difference in the 
estimation accuracy, while choosing a 
different collection of basis functions
$\{G_j\}$ gives a very effective set of
control variates. Specifically, appealing
to the first rule of thumb stated earlier,
we define two simple basis functions $G_1,G_2$
such that, a linear combination of $G_1$ and
$G_2$ can be seen, heuristically at least,
to provide a potentially accurate approximation
to the solution of the associated Poisson equation.
With these control variates,
the resulting estimation variance is reduced
by a factor ranging approximately between
15 and 55.

\medskip

\noindent
{\em General statistics of interest. }
The above general methodology 
was based on the assumption
that the goal was to estimate the posterior
mean of one of the parameters. But in many
cases it may be a different statistic that
is of interest. For example, it may be
desirable to estimate the probability that 
one of the model parameters, $\phi$ say, would
lie in a particular range, e.g., 
$\phi\in(a,b)$; here, the function
$F$ whose posterior mean is to be estimated
would be the indicator function of 
the event $\{\phi\in(a,b)\}$.
In such cases, there is little reason to believe
that the co-ordinate functions $\{G_j\}$ would
still lead to effective control variates.
Instead, we argue that it is more appropriate,
in accordance to the second rule of thumb 
derived in Section~\ref{s:cv2},
to choose the $G_j$ so that the associated
control variates $U_j:=G_j-PG_j$ are highly
correlated with~$F$.

A particular such instance is illustrated in
Example~6, where we examine a
Bayesian model-selection problem
arising from a two-threshold autoregressive model 
after all other parameters have been integrated out. 
The resulting (discrete) model 
space is sampled by a discrete
random-walk Metropolis-Hastings algorithm.
There, the quantity of interest is the posterior
probability of a particular model.
Defining $F$ as the indicator function of that 
model, a natural choice for the basis functions 
$\{G_j\}$ is to take $G_1=F$, and $G_2,G_3$
to be the indicator functions of two different
models that are close to the one whose posterior
probability is being estimated. 
Because of the discrete nature of the sampler,
the conditional expectations $PG_j$ can be 
easily evaluated, and with these
basis functions we find that the variance
of the modified estimators in (\ref{eq:Imod})
is at least 30 times smaller than that of
the standard 
ergodic averages~(\ref{eq:Iergodic}).

\subsection{Further extensions and results}
\label{s:Ifurther}

In Section~\ref{s:issues} we first briefly discuss 
two other consistent estimators for the optimal
coefficient vector $\{\theta_j^*\}$. 
One is a modified version of our earlier
estimator $\hat{\theta}_{n,\K}$ derived in
Section~\ref{s:thetastar},
and the other one was recently developed by 
\citet{meyn:book} based on the so-called
``temporal difference learning'' algorithm.
Then in Section~\ref{s:batch}
we examine the most common estimator for 
the optimal coefficient vector $\{\theta^*\}$ 
that has been used in the literature, which as
mentioned earlier is based on the method 
of batch-means. In Proposition~3
we show that the resulting estimator for $\pi(F)$ 
is typically strictly suboptimal, and that the 
amount by which its variance is {\em larger} than
the variance of our modified estimators $\mu_{n,\K}(F)$
is potentially unbounded. Moreover, the batch-means
estimator is computationally more expensive and 
generally rather ineffective, often severely so. 
This is illustrated by re-visiting the three 
MCMC examples of Section~\ref{s:simple} and 
comparing the performance of the batch-means
estimator with that of the simple ergodic
averages~(\ref{eq:Iergodic}) and of
our modified estimator $\mu_{n,\K}(F)$ in~(\ref{eq:Imod}).

Section~\ref{s:theory} provides complete theoretical 
justifications of the asymptotic arguments 
in Sections~\ref{s:cv},~\ref{s:thetastar}
and~\ref{s:issues}. Finally
we conclude with a short
summary of our results and a brief
discussion of possible further extensions 
in Section~\ref{s:conclude},
with particular emphasis on implementation
issues and on the difficulties of applying
the present methodology to general Metropolis-
Hastings samplers.

\bigskip
 
We close this introduction with a few more 
remarks on previous related work.
As mentioned earlier, \citet{henderson:phd}
takes a different path toward optimizing
the use of control variates for Markov chain
samplers. Considering primarily continuous-time 
processes, an approximation for the 
solution to the associated Poisson equation is 
derived from the so-called ``heavy traffic'' 
or ``fluid model'' approximations of the 
original process. The motivation and application
of this method is mostly related to examples
from stochastic network theory and queueing theory.
Closely related approaches are presented
by \citet{henderson-glynn:02} and
\citet{HMT:03}, where the effectiveness of
multiclass network control policies 
is evaluated via Markovian simulation.
Control variates are used for variance 
reduction, and the optimal coefficients $\{\theta_j^*\}$
are estimated via an adaptive, stochastic gradient algorithm. 
General convergence properties of ergodic estimators
using control variates are derived by \citet{henderson-simon:04},
in the case when the solution to the Poisson equation 
(either for the original chain or for an approximating
chain) is known explicitly.
\citet{kim-henderson:07} introduce two related 
adaptive methods
for tuning non-linear versions of the coefficients
$\{\theta_j\}$, when using families of control
variates that naturally admit a non-linear parameterization.
They derive asymptotic properties for these
estimators and present numerical simulation
results.

When the control variate $U=G-PG$ is
defined in terms of a function $G$ that can 
be taken as a Lyapunov function for the 
chain $\{X_n\}$, \citet{meyn:06} derives 
precise exponential asymptotics for the
associated modified estimators.
Also, \citet[Chapter~11]{meyn:book} gives 
a development of the general control variates 
methodology for Markov chain data that
parallels certain parts of our presentation
in Section~\ref{s:cv}, and discusses numerous 
related asymptotic results and implementation 
issues.

In a different direction,
\citet{stein-diaconis-et-al} draw a
connection between the use of control variates 
in MCMC and the ``exchangeable pairs'' construction 
used in Stein's method for distribution approximation.
They consider a natural class of functions as 
their control variates, and they estimate the
associated coefficients $\{\theta_j\}$ by
a simple version of the batch-means method
described in Section~\ref{s:batch}.
Finally, the recent work by
\citet{delmas-jourdain} examines a
particular case of Henderson's construction
of a control variate in the context of 
Metropolis-Hastings sampling. 
Like \citet{Hammer08}, the authors expand
the state space to include the proposals
and they first take $G=F$ and $\theta=1$
(which, in part, explains why their WR algorithm
is sometimes worse than plain Metropolis sampling).
They identify the solution of the
Poisson equation as the optimal choice for a basis
function and they seek analytical approximations. 
Then a general linear coefficient $\theta$ is 
introduced, and for a particular version of the 
Metropolis algorithm the optimal value $\theta^*$
is identified analytically.

\section{Control Variates for Markov Chains}
\label{s:cv}

\subsection{The setting}
\label{s:general}

Suppose $\{X_n\}$ is a discrete-time Markov chain with initial state
$X_0=x$, taking values in the state space $\state$, 
equipped with a $\sigma$-algebra $\clB$. In typical applications,
$\state$ will often be a (Borel measurable) subset of $\RL^d$
together the collection $\clB$ of all its (Borel) measurable
subsets. [Precise definitions and detailed assumptions are
given in \Section{theory}.] The distribution of $\{X_n\}$ is
described by its transition kernel, $P(x,dy)$,
\be
P(x,A):=\Pr\{X_{k+1}\in A\,|\,X_k=x\},
\;\;\;\;x\in\state,\;A\in\clB. 
\label{eq:kernel} 
\ee

As is well-known,
in many applications where
it is desirable to compute the expectation
$E_\pi(F):=\pi(F):=\int F\,d\pi$
of some function $F:\state\to\RL$ with respect to some
probability measure $\pi$ on $(\state,\clB)$,
although the direct computation
of $\pi(F)$ is impossible and we cannot even produce
samples from $\pi$, it is possible to
construct an
easy-to-simulate Markov chain $\{X_n\}$
which has $\pi$ as its unique invariant measure.
Under appropriate conditions, the distribution
of $X_n$ converges to $\pi$, a fact which
can be made
precise in several ways. For example,
writing
$PF$ for the function,
$$PF(x):=E_x[F(X_1)]:=E[F(X_1)\,|\,X_0=x],
\;\;\;\;x\in\state,$$
then, for any initial state $x$,
$$P^nF(x):=E[F(X_n)\,|\,X_0=x]\to\pi(F),\;\;\;\;\mbox{as}\;n\to\infty,$$
for an appropriate class of functions $F:\state\to\RL$. Furthermore,
the rate of this convergence can be quantified by the function, \be
\hat{F}(x)=\sum_{n=0}^\infty \Big [P^nF(x)-\pi(F)\Big],
\label{eq:sumFhat} 
\ee 
where $\hat{F}$ is easily seen to satisfy the
{\em Poisson equation} for $F$, namely, 
\be 
P\hat{F}-\hat{F}=-F+\pi(F).
\label{eq:poisson} 
\ee 
[To see this, at least formally, apply
$P$ to both sides of (\ref{eq:sumFhat}) and note that the resulting
series for $P\hat{F}-\hat{F}$ becomes telescoping.]

The above results describe how the distribution
of $X_n$ converges to $\pi$. In terms of estimation,
the quantities of interest are the ergodic averages,
\be
\mu_n(F):=\frac{1}{n}\sum_{i=0}^{n-1}F(X_i).
\label{eq:ergodicAv}
\ee
Again, under appropriate conditions the ergodic theorem
holds,
\be
\mu_n(F)\to\pi(F),\;\;\;\;\mbox{a.s., as}\;n\to\infty,
\label{eq:ergodic}
\ee
for an appropriate class of functions $F$.
Moreover, the rate of this convergence is quantified
by an associated central limit theorem, which states
that,
$$\sqrt{n}[\mu_n(F)-\pi(F)]=
\frac{1}{\sqrt{n}}\sum_{i=0}^{n-1}[F(X_i)-\pi(F)] \weakly
N(0,\sigma_F^2), \;\;\;\;\mbox{as}\;n\to\infty,$$ where
$\sigma_F^2$, the asymptotic variance of $F$, is given by,
$\sigma_F^2:=\lim_{n\to\infty}\VAR_\pi(\sqrt{n}\mu_n(F))$.
Alternatively, it can be expressed in terms of
$\hat{F}$ as,
\be
\sigma_F^2=\pi\Big(\hat{F}^2-(P\hat{F})^2\Big).
\label{eq:varF}
\ee

The results in equations (\ref{eq:sumFhat})
and (\ref{eq:varF}) clearly indicate that
it is
useful to be able to compute the solution
$\hat{F}$ to the Poisson equation for $F$.
In general this is a highly nontrivial task,
and, for chains arising in typical applications,
it is impossible for all practical purposes; 
see, e.g., the relevant
comments in \citet{henderson:phd} and \citet{meyn:book}.
Nevertheless, 
the function $\hat{F}$
will play a central role throughout our
subsequent development.

\subsection{Control variates}
\label{s:cv2}

Suppose that, for some Markov chain
$\{X_n\}$ with transition kernel $P$ and invariant
measure $\pi$, the ergodic averages
$\mu_n(F)$ as in (\ref{eq:ergodicAv}) 
are used
to estimate the mean $\pi(F)=\int F\,d\pi$
of some function $F$
under $\pi$. In many applications,
although the estimates $\mu_n(F)$
converge to $\pi(F)$ as $n\to\infty$,
the associated asymptotic variance $\sigma_F^2$
is large and the
convergence is very slow.

In order to reduce the variance, we employ the idea of using control
variates, as in the case of simple Monte Carlo with 
independent and identically distributed ($\iid$) samples;
see, for example, the standard texts of
\citet{robert:book}, \citet{liu:book}, \citet{givens:book}, 
or the paper by
\citet{glynn-szechtman} for extensive discussions.
Given one or more functions $U_1,U_2,\ldots,U_k$,
the {\em control variates},
such that $U_j:\state\to\RL$ 
and $\pi(U_j)=0$ for all $j=1,2,\ldots,k$,
let $\theta=(\theta_1,\theta_2,\ldots,\theta_k)^t$
be an arbitrary, constant vector in $\RL^k$,
and define,
\be 
F_\theta:=F-\langle\theta, U\rangle
=F-\sum_{j=1}^k\theta_jU_j, 
\label{eq:Ftheta} 
\ee 
where 
$U:\state\to\RL^k$ denotes the column vector,
$U=(U_1,U_2,\ldots,U_k)^t$.
[Here and throughout the paper
all vectors are column vectors unless
explicitly stated otherwise, and
$\langle\cdot,\cdot\rangle$
denotes the usual Euclidean inner product.]

We consider the
modified estimators,
\be
\mu_n(F_\theta)
=\mu_n(F)-\langle\theta,\mu_n(U)\rangle
=\mu_n(F)-\sum_{j=1}^k\theta_j\,\mu_n(U_j),
\label{eq:modified}
\ee
for $\pi(F)$.
The ergodic theorem 
(\ref{eq:ergodic}) guarantees
that the estimators $\{\mu_n(F_\theta)\}$ 
are consistent with probability
one, and it is natural to seek particular choices
for $U$ and $\theta$ so that the asymptotic
variance $\sigma_{F_\theta}^2$ of the modified
estimators
is significantly smaller than the variance $\sigma_F^2$
of the standard ergodic averages $\mu_n(F)$.

Throughout this work, 
we will concentrate exclusively on 
the following class of control variates $U$ 
proposed by \citet{henderson:phd}. 
For arbitrary ($\pi$-integrable) functions
$G_j:\state\to\RL$
define,
$$U_j:=G_j-PG_j,
\;\;\;\;j=1,2,\ldots,k.$$
Then the invariance of $\pi$ under $P$ and  the integrability
of $G_j$ guarantee that $\pi(U_j)=0$.

In the remainder of this section we derive
some simple, general guidelines for choosing functions
$\{G_j\}$ that produce effective control 
variates $\{U_j\}$. This issue is revisited
in more detail in the Bayesian MCMC
context in Section~\ref{s:basic}.

Suppose, at first, that we have complete freedom
in the choice of the functions $\{G_j\}$, so that 
we may take $k=1$, a single $U=G-PG$, and $\theta=1$
without loss of generality. Then the goal is
to make the asymptotic variance of $F-U=F-G+PG$
as small as possible. But, in view of the Poisson equation
(\ref{eq:poisson}), we see that the choice $G=\hat{F}$ yields, 
\ben
F-U=F-\hat{F}+P\hat{F}=\pi(F), 
\een
which has zero
variance. Therefore, the general principle
for selecting a single function $G$ is:

\begin{center}
{\em Choose a control variate $U=G-PG$ with
$G\approx\hat{F}$}.
\end{center}

\noindent
As mentioned above, it is typically impossible
to compute $\hat{F}$ for realistic models used in
applications. But it is often possible to come
up with a guess $G$ that approximates $\hat{F}$,
or at least with a collection of functions $\{G_j\}$
such that $\hat{F}$ can be approximated 
as linear combination of the $\{G_j\}$.
Thus, our first concrete rule of thumb 
for choosing $\{G_j\}$ states:

\begin{center}
{\em Choose control variates $U_j=G_j-PG_j$, $j=1,2,\ldots,k$
with respect to a collection\\
of {\em\bf basis functions} $\{G_j\}$,
such that $\hat{F}$ can be approximately expressed\\
as a linear combination of the $\{G_j\}$.}
\end{center}

\noindent
The terminology 
{\em basis functions} for the $\{G_j\}$ is meant
to emphasize the fact that, although $\hat{F}$ is
not known, it is expected that it can be 
approximately expressed 
in terms of the $\{G_j\}$
via a linear expansion
of the form $\hat{F}\approx\sum_{j=1}^k\theta_jG_j$.

Once the basis functions $\{G_j\}$ are selected,
we form the modified estimators $\mu_n(F_\theta)$
with respect to the function $F_\theta$ as in
(\ref{eq:Ftheta}),
$$F_\theta=F-\langle\theta,U\rangle
=F-\langle\theta,G\rangle+\langle\theta,PG\rangle,$$
where, for a vector of functions $G=(G_1,G_2,\ldots,G_k)^t$,
we write $PG$ for the corresponding vector,
$(PG_1,PG_2,\ldots,PG_k)^t$.
The next task is to choose $\theta$ so that the
resulting variance,
$$\sigma_\theta^2:=\sigma_{F_\theta}^2=
\pi\Big(\hat{F}_\theta^2-(P\hat{F}_\theta)^2\Big),$$
is minimized. Note that, from the definitions,
\ben
\hat{U_j}=G_j\;\;\mbox{for each $j$},
\;\;\;\;\;\mbox{and}
\;\;\;\;
\hat{F}_\theta=\hat{F}-\langle\theta,G\rangle.
\een
Therefore, by (\ref{eq:varF}) and linearity,
\be
\sigma_\theta^2
&=&
\sigma_F^2-2\pi\Big(
\hat{F}\langle\theta,G\rangle-P\hat{F}\langle\theta,PG\rangle
\Big)
+\pi\Big(
\langle\theta,G\rangle^2-
\langle\theta,PG\rangle^2
\Big).
\label{eq:varTheta}
\ee

To find the optimal $\theta^*$ which minimizes
the variance $\sigma_\theta^2$, differentiating
the quadratic
$\sigma_\theta^2$ with respect to each $\theta_j$
and setting the derivative equal to zero, yields,
in matrix notation,
$$\Gamma(G)\theta^*=\pi(\hat{F}G-(P\hat{F})(PG)),$$
where the $k\times k$ matrix $\Gamma(G)$ has entries,
$\Gamma(G)_{ij}=\pi(G_iG_j-(PG_i)(PG_j))$. Therefore,
\be
\theta^*=
\Gamma(G)^{-1}\pi(\hat{F}G-(P\hat{F})(PG)),
\label{eq:thetaStar-k}
\ee
as long as $\Gamma(G)$ is invertible.
Once again, this expression depends on $\hat{F}$,
so it is not immediately clear how to estimate
$\theta^*$ directly from the data $\{X_n\}$.
The issue of estimating the optimal
coefficient vector $\theta^*$ is
addressed in detail in Section~\ref{s:thetastar};
but first let us interpret $\theta^*$. 

For simplicity, consider again the case of
a single control variate $U=G-PG$ based on 
a single function $G$. Then the value of
$\theta^*$ in (\ref{eq:thetaStar-k}) simplifies to,
\be
\theta^*
=\frac
{\pi\big(\hat{F}G-(P\hat{F})(PG)\big)}
{\pi(G^2-(PG)^2)}
=\frac
{\pi\big(\hat{F}G-(P\hat{F})(PG)\big)}
{\sigma_U^2},
\label{eq:thetaStar}
\ee
where the second equality 
follows from the earlier 
observation that $\hat{U}=G$.
Alternatively,
starting from the expression,
$\sigma_\theta^2=\lim_{n\to\infty}
\VAR_\pi(\sqrt{n}\mu_n(F_\theta)),$
simple calculations lead to,
\be
\sigma_\theta^2=\sigma_F^2+\theta^2\sigma_U^2
-2\theta\sum_{n=-\infty}^\infty\COV_\pi(F(X_0),U(X_n)),
\label{eq:varseries}
\ee
so that $\theta^*$ can also be expressed as,
\be
\theta^*=
\frac{1}{\sigma_U^2}
\sum_{n=-\infty}^\infty\COV_\pi(F(X_0),U(X_n)),
\label{eq:thetaRatio}
\ee
where $\COV_\pi$ denotes the covariance 
for the stationary version of the chain,
i.e., since $\pi(U)=0$, we have
$\COV_\pi(F(X_0),U(X_n))=E_\pi[F(X_0)U(X_n)]$,
where $X_0\sim\pi$. Then~(\ref{eq:thetaRatio})
leads to the optimal asymptotic variance,
\be
\sigma_{\theta^*}^2
=\sigma_F^2 -\frac{1}{\sigma_U^2}
\Big[\sum_{n=-\infty}^\infty\COV_\pi(F(X_0),U(X_n))\Big]^2.
\label{eq:thetaseries}
\ee
Therefore, in order to reduce the variance, it is
desirable that  
the correlation between $F$ and $U$ be
as large as possible. This leads to our second
rule of thumb for selecting basis functions:

\begin{center}
{\em Choose control variates $U=G-PG$ so
that each $U_j$ is highly correlated with $F$.}
\end{center}

\noindent
Incidentally, note that,
comparing the expressions
for $\theta^*$ in (\ref{eq:thetaStar}) and
(\ref{eq:thetaRatio}) implies that,
\be
\sum_{n=-\infty}^\infty\COV_\pi(F(X_0),U(X_n))
=
    \pi\big(\hat{F}G-(P\hat{F})(PG)\big).
\label{eq:cov}
\ee

\section{Estimating the Optimal Coefficient Vector $\theta^*$}
\label{s:thetastar}

Consider, as before, the problem of estimating
the mean $\pi(F)$ of a function $F:\state\to\RL$
based on samples from an ergodic Markov chain 
$\{X_n\}$ with unique invariant measure $\pi$
and transition kernel $P$. Instead of using the
ergodic averages $\mu_n(F)$ as in (\ref{eq:ergodicAv}),
we select a collection of {\em basis functions}
$\{G_j\}$ and form the control variates 
$U_j=G_j-PG_j$, $j=1,2,\ldots,k$. The mean
$\pi(F)$ is then estimated by the modified
estimators $\mu_n(F_\theta)$ as in (\ref{eq:modified}),
for a given coefficient vector $\theta\in\RL^k$.

Under the additional assumption of reversibility,
in this section we introduce a consistent procedure 
for estimating the optimal coefficient
vector $\theta^*$ based on the same sample $\{X_n\}$.
Then in Section~\ref{s:basic} we give more detailed 
guidelines for choosing the $\{G_j\}$.

Recall that, once the basis
functions $\{G_j\}$ have been selected,
the optimal coefficient vector $\theta^*$ was
expressed in (\ref{eq:thetaStar-k}) as,
$\theta^*=\Gamma(G)^{-1}\pi(\hat{F}G-(P\hat{F})(PG)),$
where $\Gamma(G)_{ij}=\pi(G_iG_j-(PG_i)(PG_j))$, 
$1\leq i,j\leq k$.
But, in view of equation (\ref{eq:cov})
derived above, the entries $\Gamma(G)_{ij}$
can also be written,
\be
\Gamma(G)_{ij}
&:=&
	\pi(G_iG_j-(PG_i)(PG_j))
	\label{eq:Gamma}
	\\
&=&
	\pi(\hat{U_i}G_j-(P\hat{U}_i)(PG_j))
	\;=\;\sum_{n=-\infty}^\infty\COV_\pi(U_i(X_0),G_j(X_n)).
	\nonumber
\ee
This indicates that $\Gamma(G)$ has the structure
of a covariance matrix and, in particular,
it suggests that $\Gamma(G)$ should be positive
semidefinite. Indeed:

\medskip

\noindent
{\bf Proposition 1. } {\em Let $\K(G)$ denote the covariance
matrix of the random variables
$$Y_j:=G_j(X_1)-PG_j(X_0),
\;\;\;\;
j=1,2,\ldots,k,$$
where $X_0\sim \pi$. Then $\Gamma(G)=\K(G)$,
that is,
for all $1\leq i,j\leq k$,
\be
\pi(G_iG_j-(PG_i)(PG_j)) =
	\K(G)_{ij}:=
    E_\pi\Big[
    \Big(G_i(X_1)-PG_i(X_0)\Big)
    \Big(G_j(X_1)-PG_j(X_0)\Big)
    \Big].
\label{eq:lemma2}
\ee
}

\noindent
{\sc Proof. }
Expanding the right-hand side of (\ref{eq:lemma2})
yields,
$$  \pi(G_iG_j)
    -E_\pi[G_i(X_1)PG_j(X_0)]
    -E_\pi[G_j(X_1)PG_i(X_0)]
    +\pi((PG_i)(PG_j)),
$$
and the result follows upon noting that the
second and third terms above are both equal to
the fourth. To see this, observe that the
second term can be rewritten as,
\ben
    E_\pi\Big\{E\Big[G_i(X_1)PG_j(X_0)\,\Big|\,X_0\Big]\Big\}
=
    E_\pi\Big[E[G_i(X_1)\,|\,X_0]PG_j(X_0)\Big]
=
    \pi((PG_i)(PG_j)),
\een
and similarly for the third term.
\qed

Therefore, using Proposition~1 the optimal coefficient 
vector $\theta^*$ can also be expressed as,
\be
\theta^*=
\K(G)^{-1}\pi(\hat{F}G-(P\hat{F})(PG)).
\label{eq:thetaStar-k_2}
\ee

Now assume that the chain $\{X_n\}$ is reversible.
Writing $\Delta=P-I$ for the {\em generator} of 
$\{X_n\}$, reversibility is equivalent to the
statement that $\Delta$ is self-adjoint as a
linear operator on the space $L_2(\pi)$. 
In other words,
$$\pi(F\,\Delta G)=\pi(\Delta F\, G),$$
for any two functions $F,G\in L_2(\pi)$. 
Our first main theoretical result 
is that the optimal coefficient vector 
$\theta^*$ admits a representation that does
not involve the solution $\hat{F}$ of the
associated Poisson
equation for $F$:

\medskip

\noindent
{\bf Proposition 2. } {\em If the chain $\{X_n\}$ is
reversible, then the optimal coefficient vector
$\theta^*$ for the control variates
$U_i=G_i-PG_i$, $i=1,2,\ldots,k$ can be expressed
as,
\be
\theta^*=
\theta^*_\trev=\Gamma(G)^{-1}
\pi\big((F-\pi(F))(G+PG)\big),
\label{eq:theta_rev_k}
\ee
or, alternatively,
\be
\theta^*_\trev=\K(G)^{-1}
\pi\big((F-\pi(F))(G+PG)\big),
\label{eq:theta_rev_k_2}
\ee
where the matrices $\Gamma(G)$ and $\K(G)$
are defined in~{\em (\ref{eq:Gamma})} and~{\em (\ref{eq:lemma2})},
respectively.}

\medskip

\noindent
{\sc Proof. }
Let $\bar{F}=F-\pi(F)$ denote the centered
version of $F$, and recall that $\hat{F}$ solves
Poisson's equation for $F$, so $P\hat{F}=\hat{F}-\bar{F}$.
Therefore, using the fact that $\Delta$
is self-adjoint on each component of $G$,
\ben
\pi\big(\hat{F}G-(P\hat{F})(PG)\big)
&=&
\pi\big(\hat{F}G-(\hat{F}-\bar{F})(PG)\big)\\
&=&
\pi\big(\bar{F}PG-\hat{F}\Delta G\big)\\
&=&
\pi\big(\bar{F}PG-\Delta\hat{F} G\big)\\
&=&
\pi\big(\bar{F}PG+\bar{F} G\big)\\
&=&
\pi\big(\bar{F}(G+PG)\big).
\een
Combining this with
(\ref{eq:thetaStar-k})
and 
(\ref{eq:thetaStar-k_2}),
respectively, proves the two claims
of the proposition.
\qed

The expression
(\ref{eq:theta_rev_k_2})
suggests estimating
$\theta^*$ via,
\be
\hat{\theta}_{n,\K}
&=&
    \K_n(G)^{-1}
    [\mu_n(F(G+PG))-
    \mu_n(F)\mu_n(G+PG)],
\label{eq:theta-n-K}
\ee
where the empirical $k\times k$
matrix $\K_n(G)$ is
defined by,
\be
(\K_n(G))_{ij}
&=&
\frac{1}{n-1}\sum_{t=1}^{n-1}
(G_i(X_t)-PG_i(X_{t-1}))
(G_j(X_t)-PG_j(X_{t-1})).
\label{eq:theta-optimal-est}
\ee
The resulting estimator
$\mu_n(F_{\hat{\theta}_{n,\K}})$
for $\pi(F)$
based on the vector of
control variates $U=G-PG$
and the estimated coefficients
$\hat{\theta}_{n,\K}$
is defined as,
\be
\mu_{n,\K}(F)
:=
    \mu_n(F_{\hat{\theta}_{n,\K}})
    \;=\;\mu_n(F)-\langle\hat{\theta}_{n,\K},\mu_n(U)\rangle.
    \label{eq:multi-est2}
\ee
This will be the main estimator used in the remainder
of the paper.

\section{The Choice of Basis Functions and the Basic Methodology}
\label{s:basic}

Given an ergodic, reversible Markov chain $\{X_n\}$ 
with invariant measure $\pi$, and a function $F$ 
whose mean $\pi(F)$ is to be estimated based on samples
from the chain, we wish to find a vector of control 
variates $U=G-PG$ so that,
for an appropriately chosen coefficient vector $\theta$,
the variance of the modified 
estimates $\mu_n(F_\theta)=\mu_n(F)-\langle\theta,U\rangle$
in (\ref{eq:Ftheta}) is significantly lower than that
of the simple ergodic averages $\mu_n(F)$ 
in~(\ref{eq:ergodicAv}).

As noted in the Introduction, like with simple 
Monte Carlo estimation based on i.i.d.\ samples, 
there cannot possibly be 
a single, universal family of control variates
that works well in every problem. Indeed,
the choice 
of effective basis functions $G=\{G_j\}$ 
is important for constructing useful control 
variates $U_j=G_j-PG_j$ for variance reduction.
Nevertheless -- and perhaps somewhat surprisingly --
as shown next, it {\em is} possible to select 
basis functions $\{G_j\}$ that provide very 
effective variance reduction
in a wide range of MCMC estimation scenarios
stemming from Bayesian inference problems,
primarily those where inference is performed
via a conjugate random-scan Gibbs sampler.
Examples of this basic methodology 
are presented in Section~\ref{s:simple}.
Extensions and further examples 
are discussed in Section~\ref{s:generalMeth}.

Our starting point is the observation that
in order to apply the results developed so far, 
the MCMC sampler at hand needs to be reversible 
so that the estimator $\hat{\theta}_{n,\K}$ 
in (\ref{eq:theta-n-K}) for the optimal 
coefficient vector $\theta^*$ can be used,
and also it is necessary that 
the one-step expectations 
$PG_j(x)=E[G_j(X_{n+1})|X_n=x]$ of the basis
functions $G_j$ should be explicitly 
computable in closed form.
Probably the most natural, general family 
of MCMC algorithms that satisfy these two
requirements is the collection of conjugate 
random-scan Gibbs samplers, with a target 
distribution $\pi$ arising as the posterior 
density of the parameters in a Bayesian 
inference study.
Moreover, since for large sample sizes the
posterior is approximately normal --
see, e.g.,
\citet{bickel-yahav:69},
\citet{blackwell:TR},
\citet{bunke-milhaud:98},
\citet{ibragimov:book} -- we focus on 
the case where $\pi$ is a multivariate 
normal distribution.

According to the discussion
in Section~\ref{s:cv2}, the main goal in choosing 
the basis functions $G=\{G_j\}$ is that
it should be possible to effectively approximate
the solution $\hat{F}$ of the Poisson equation
for $F$
as a linear combination of the $G_j$, i.e.,
$\hat{F}\approx \sum_{j=1}^k\theta_j G_j$.
It is somewhat remarkable that,
in the case of random-scan Gibbs sampler
with a Gaussian target density, the Poisson 
equation can be solved explicitly, 
and its solution is of a particularly 
simple form:

\medskip

\noindent
{\bf Theorem 1. } {\em Let $\{X_n\}$ denote the Markov
chain constructed from the random-scan Gibbs sampler
used to simulate from an arbitrary (nondegenerate)
multivariate normal distribution
$\pi\sim N(\mu,\Sigma)$ in $\RL^k$.
If the goal is to estimate the mean of the first
component of $\pi$, then letting $F(x)=x^{(1)}$ for 
$x=(x^{(1)},x^{(2)},\ldots,x^{(k)})^t\in\RL^k,$
the solution $\hat{F}$ of the Poisson equation
for $F$ can be expressed as linear combination
of the co-ordinate basis functions $G_j(x):=x^{(j)}$,
$x\in\RL^k$, $1\leq j\leq k$,
\be
\hat{F}=\sum_{j=1}^k\theta_j G_j.
\label{eq:solution}
\ee
Moreover, writing $Q=\Sigma^{-1}$,
the coefficient vector $\theta$ in~{\em (\ref{eq:solution})}
is given by the first row of the matrix
$k(I-A)^{-1}$,
where $A$ has entries $A_{ij}=-Q_{ij}/Q_{ii}$,
$1\leq i\neq j\leq k$, $A_{ii}=0$ for all $i$,
and $(I-A)$ is always invertible.}

\medskip


\noindent
{\sc Proof. }
Let $H$ denote the candidate solution to the Poisson
equation, $H(x)=\sum_j\theta_jx^{(j)}$, and write 
$X=(X^{(1)},X^{(2)},\ldots,X^{(k)})$
for a random vector with distribution $\pi\sim N(\mu,\Sigma)$. 
Since $\pi$ is nondegenerate,
$\Sigma$ is nonsingular and so
the precision matrix $Q$ exists
and its diagonal entries are nonzero.
Since the conditional expectation of a component $X^{(j)}$
of $X$ given the values of the remaining 
$X^{(-j)}:=(X^{(1)},\ldots,X^{(j-1)},X^{(j+1)},\ldots,X^{(k)})$
is $\mu^{(j)}+ \sum_\ell A_{j\ell}\,[X^{(\ell)}-\mu^{(\ell)}]$, 
we have,
$$PH(x)=\sum_j\theta_j\left\{\frac{k-1}{k}x^{(j)}+\frac{1}{k}
\Big[\mu^{(j)}+
\sum_\ell A_{j\ell}\Big(x^{(\ell)}-\mu^{(\ell)}\Big)
\Big]\right\},$$
so that,
\ben
PH(x)-H(x)
&=&
-\frac{1}{k}\sum_j\theta_j\left\{
(x^{(j)}-\mu^{(j)})-\sum_\ell A_{j\ell}(x^{(\ell)}-\mu^{(\ell)})
\right\}\\
&=&
-\frac{1}{k}\theta^t(I-A)(x-\mu),
\een
where we have used the fact that 
$\sum_{\ell}A_{j\ell}(x^{(j)}-\mu^{(j)})
=\sum_{\ell\neq j}A_{j\ell}(x^{(j)}-\mu^{(j)})$,
since the diagonal entries of $A$ are all zero.
In order for this to be equal to $-F(x)+\pi(F)=-(x^{(1)}-\mu^{(1)})$
for all $x$, it suffices to choose $\theta$ such that
$\theta^t(I-A)=(k,0,\ldots,0)$, as claimed. Finally,
to see that $(I-A)$ is nonsingular (and hence invertible),
note that its determinant is equal to 
$[\prod_j(1/Q_{jj})]\det(Q)$, which is nonzero
since $\Sigma$ is nonsingular by assumption. 
\qed


In terms of MCMC estimation, the statement 
of Theorem~1 can be rephrased as follows:
Suppose that samples from a multivariate 
Gaussian distribution are simulated via
a random-scan Gibbs sampler in order to estimate the
mean of one of its components. Then, using the linear
basis functions $G_j(x)=x^{(j)}$ to construct
a vector of control variates $U=G-PG$,
the modified estimator $\mu_{n,\K}(F)$ as 
in (\ref{eq:multi-est2}) should have dramatically
smaller variance than the standard ergodic
averages $\mu_n(F)$; in fact, the asymptotic
variance of $\mu_{n,\K}(F)$ in the central limit theorem
is equal to {\em zero}. This theoretical prediction
is verified in a simple simulation example presented 
in Section~\ref{s:simple}.

Thus motivated, we now state the basic version of the variance
reduction methodology proposed in this work.

\begin{center}
\framebox{
\shortstack[l]{
\smallskip\\
\hspace{1.2in}
{\sc Outline of the Basic Methodology}\\
\smallskip\\
\noindent
$(i)$ Given:\\
\hspace{0.2in}
$\bullet\;$
A multivariate posterior distribution 
$\pi(x)=\pi(x^{(1)},x^{(2)},\ldots,x^{(d)}))\;\;$\\
\hspace{0.2in}
$\bullet\;$
A reversible Markov chain $\{X_n\}$ with stationary
distribution $\pi$\\ 
\hspace{0.2in}
$\bullet\;$
A sample of length $n$ from the chain $\{X_n\}$
\smallskip\\
$(ii)$ Goal:\\
\hspace{0.2in}
$\bullet\;$
Estimate the posterior mean $\mu^{(i)}$ of $x^{(i)}$\\
\smallskip\\
$(iii)$ Define:\\
\hspace{0.2in}
$\bullet\;$
$F(x)=x^{(i)}$\\
\hspace{0.2in}
$\bullet\;$
Basis functions as the co-ordinate functions $G_j(x)=x^{(j)}$\\
\hspace{0.5in} 
for all $j$
for which
$PG_j(x):=E[X_{n+1}^{(j)}\,|\,X_n=x]$\\
\hspace{0.5in} 
is computable in closed form\\
\hspace{0.2in}
$\bullet\;$
The corresponding control variates 
$U_j=G_j-PG_j$\\
\smallskip\\
$(iv)$ Estimate:\\
\hspace{0.2in}
$\bullet\;$
The optimal coefficient vector $\theta^*$ by
$\hat{\theta}_{n,\K}$ as in (\ref{eq:theta-n-K})\\
\hspace{0.2in}
$\bullet\;$
The quantity of interest $\mu^{(i)}$ by the modified
estimators\\
\hspace{0.5in}
$\mu_{n,\K}(F)$ as in (\ref{eq:multi-est2})
\medskip
}
}
\end{center}

Before examining the performance of this methodology
in practice we recall that the Markov chain in Theorem~1 
is perhaps the single most complex example 
of a Markov chain naturally arising in an applied context,
for which there is an explicit 
solution to the Poisson equation. 

Finally we mention that, even if the posterior 
distribution $\pi$ to be simulated is not Gaussian, 
in many cases it is possible to re-parametrize
the model so that $\pi$ is either Gaussian
or approximately Gaussian; see, e.g.,
\citet{hilsmi93} and \citet{tibshirani:94}.


\section{MCMC Examples of the Basic Methodology}
\label{s:simple}

Here we present three examples 
of the application of the basic methodology
outlined in the previous section.
The examples below as well as those
in Section~\ref{s:generalMeth} are chosen
as representative cases covering a broad class 
of real applications of Bayesian inference.

In Example~1, a bivariate normal density is
simulated by random-scan Gibbs sampling.
This setting is considered primarily as an
illustration of the result of Theorem~1, and
also as a simplified version of many examples 
in the large class of inference studies with 
an approximately normal posterior distribution.
As expected, the variance of the modified 
estimators in this case is dramatically smaller.

Example~2 considers a case of Bayesian inference 
via MCMC, with a large hierarchical normal 
linear model. In part due to the complexity of 
the model, it is perhaps natural to expect that 
the posterior is approximately Gaussian, and, 
indeed, the basic methodology of the previous
section is shown to provide very significant
variance reduction.

Finally, in Example~3 the use of the basic 
methodology is illustrated in the case of 
Metropolis-within-Gibbs sampling
from the posterior of a ``difficult'' model,
where the use of heavy-tailed priors results
in heavy-tailed posterior densities. Such densities 
are commonly met in, for example, spatial statistics; 
see \citet{DelRob03} for an illustrative example.
Even in this case where the posterior is certainly
not approximately normal, it is demonstrated that the 
basic methodology is still very effective 
in reducing the estimation variance.
This example also illustrates the point that,
often, not all basis function $G_j$ can be 
easily used in the construction of control
variates, as indicated in the second part
of step~$(iii)$ of the description of the
basic methodology above.

\paragraph{Example 1. The bivariate Gaussian through 
the random-scan Gibbs sampler. }
Let $(X,Y)\sim\pi(x,y)$
be an arbitrary
bivariate normal distribution, where, without loss of generality, we
take the expected values of both $X$ and $Y$ to be zero and the
variance of $X$ to be equal to one. Let $\VAR(Y)=\tau^2$ and the
covariance $E(XY)=\rho\tau$ for some $\rho\in(-1,1)$. Given
arbitrary initial values $x_0=x$ and $y_0=y$, 
the random-scan Gibbs sampler selects one of the two
co-ordinates at random, and either
updates $y$ by sampling from $\pi(y|x)\sim N(\rho\tau
x,\,\tau^2(1-\rho^2))$, or $x$ 
from $\pi(x|y)\sim N(\frac{\rho}{\tau}
y,\,1-\rho^2)$. Continuing this way produces a reversible Markov
chain $\{(X_n,Y_n)\}$ with distribution converging to $\pi$.
To estimate the expected value of
$X$ under $\pi$ we set $F(x,y)=x$,
and define the basis functions
$G_1(x,y)=x$ and $G_2(x,y)=y$.
The corresponding functions $PG_1$ 
and $PG_2$
are easily computed as,
$$
PG_1(x,y)=\frac{1}{2}\Big[x+
\frac{\rho y}{\tau}\Big]
\;\;\;\;\mbox{and}\;\;\;\;
PG_2(x,y)=\frac{1}{2}(y+
\rho\tau x).
$$

The parameter values are chosen as,
$\rho=0.99$ and $\tau^2=10$,
so that the two components are 
highly correlated and the sampler
converges slowly,
making the variance of the 
standard estimates $\mu_n(F)$
large. Using the samples 
produced by the resulting chain
with initial values $x_0=y_0=0.5$,
we examine the performance
of the modified estimator
$\mu_{n,\K}(F)$,
and compare it with
the performance of 
the standard ergodic averages
$\mu_n(F)$.

Figure~\ref{f:example2_multi}
depicts a typical realization
of the sequence of estimates
obtained by the two estimators,
for $n=20000$ simulation steps;
the factors by which the variance of $\mu_n(F)$ 
is larger than that of $\mu_{n,\K}(F)$ 
are shown in Table~\ref{tab:ex2_multi}.
In view of Theorem~1, it is not 
surprising that the estimator 
$\mu_{n,\K}(F)$
is clearly much more effective
than $\mu_n(F)$.

\begin{figure}[ht!]
\centerline{\includegraphics[width=4.4in]{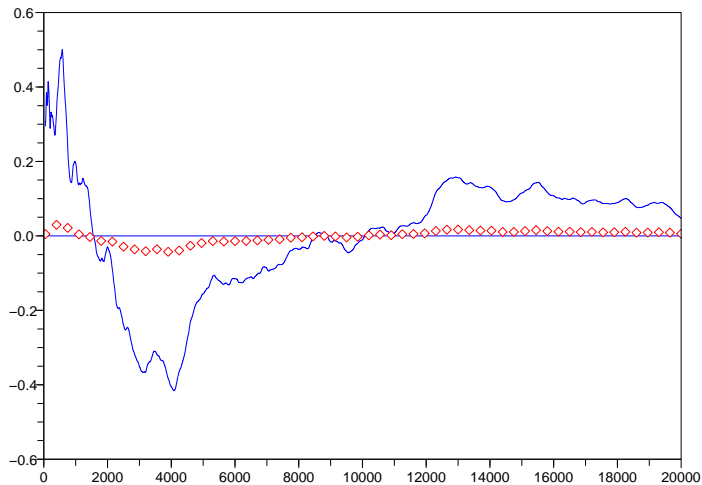}}
\caption{The sequence of the standard ergodic averages
is shown as a solid blue line
and the modified estimates
$\mu_{n,\K}(F)$
as red diamonds.
For visual clarity,
the values $\mu_{n,\K}(F)$
are plotted only every 350 simulation
steps.}
\flabel{example2_multi}
\end{figure}

\begin{table}[ht!]
  \begin{center}
    \begin{tabular}{|c||c|c|c|c|c|c|}
      \hline
      \multicolumn{7}{|c|}{\bf Variance reduction factors} \\
      \hline
      & \multicolumn{6}{|c|}{\em Simulation steps}\\
        \hline
{\em Estimator}& $n=1000$& $n=10000$& $n=50000$& $n=100000$& $n=200000$&
							$n=500000$\\
     \hline
$\mu_{n,\K}(F)$     & 4.13 & 27.91 & 122.4 & 262.5 & 445.0 & 1196.6\\
     \hline
     \end{tabular}
  \end{center}
     \caption{Estimated factors by which the
    variance of $\mu_n(F)$
    is larger than the 
    variance of
    $\mu_{n,\K}(F)$,
    after
    $n=1000,10000,50000,100000,200000$ and $500000$ 
	simulation
    steps.}
     \label{tab:ex2_multi}
\end{table}

In this and in all subsequent examples,
the reduction in the variance was computed from
independent repetitions of the same experiment:
For $\mu_n(F)$, $T=200$ different estimates
$\mu^{(i)}_n(F)$, for $i=1,2,\ldots,T$, were
obtained, and the variance of $\mu_n(F)$ 
was estimated by,
\ben
\frac{1}{T-1}\sum_{i=1}^{T}
[\mu^{(i)}_n(F)-\bar{\mu}_n(F)]^2,
\een
where
$\bar{\mu}_n(F)$ is the average of the $\mu_n^{(i)}(F)$.
The same procedure was used to estimate the
variance of $\mu_{n,\K}(F)$.

\paragraph{Example 2. A hierarchical normal linear model. }
In an early application of MCMC in Bayesian statistics, 
\citet{hills} illustrate the use of Gibbs sampling 
for inference in a large hierarchical normal linear model.  
The data consist of $N=5$ weekly
weight measurements of $\ell=30$ young rats,
whose weight is assumed to increase linearly in time,
so that,
\[
Y_{ij} \sim N \left( \alpha_i + \beta_i x_{ij},\sigma^2_c \right),
\hspace{16pt} 1\leq i\leq \ell,\;1\leq j\leq N,
\]
where the $Y_{ij}$ are the measured weights 
and the $x_{ij}$ denote the corresponding 
rats' ages (in days). The population structure 
and the conjugate prior
specification are assumed to be
of the customary normal-Wishart-inverse gamma
form: For $i=1,2,\ldots,\ell$,
\begin{eqnarray*}
\phi_i & = & \left( \begin{array}{c} \alpha_i \\
\beta_i\end{array} \right) \sim
N (\mu_c, \Sigma_c )\\
\mu_c & = & \left( \begin{array}{c} \alpha_c \\ 
\beta_c \end{array} \right) \sim N \left( \eta, C \right)  \\
\Sigma_c^{-1} & \sim &  \mbox{W}(  (\rho R )^{-1}, \rho )  \\
\sigma^2_c & \sim & \mbox{IG} \Big( \frac{\nu_0}{2} , \frac{\nu_0
\tau^2_0}{2}  \Big),
 \end{eqnarray*}
with known values for $\eta, C,\nu_0,\rho,R$ and $\tau_0$. 

The posterior $\pi$ has $k:=2\ell+2+3+1=66$ parameters, 
and MCMC samples from $((\phi_i),\mu_c,\Sigma_c,\sigma_c^2)\sim \pi$
can be generated via conjugate Gibbs sampling since
the full conditional densities of all four parameter
blocks $(\phi_i)$, $\mu_c$, $\Sigma_c$, and $\sigma^2_c$,
are easily identified explicitly in terms of standard
distributions; cf.\ \citet{hills}.
For example, 
conditional on 
$(\phi_i),\Sigma_c,\sigma_c^2$ and the observations
$(Y_{ij})$, the means $\mu_c$ have a bivariate
normal distribution with covariance matrix
$V:= (\ell\Sigma_c^{-1}+C^{-1})^{-1}$
and mean,
\be
V\Big[\Sigma_c^{-1}\Big(\sum_i\phi_i\Big)+C^{-1}\eta\Big].
\label{eq:meanhills}
\ee

Suppose, first, that we wish to estimate the posterior mean
of $\alpha_c$. We use a four-block, random-scan Gibbs 
sampler, which at each step selects one of the four 
blocks at random and replaces the current values
of the parameter(s) in that block with a draw from
the corresponding full conditional density.
We set $F((\phi_i),\mu_c,\Sigma_c,\sigma^2_c)=\alpha_c$,
and construct control variates according to the
basic methodology by first defining $k=66$ basis
functions $G_j$ and then computing the one step expectations
$PG_j$. For example, numbering each $G_j$ with the 
corresponding index in the order in which it appears
above, we have
$G_{61}((\phi_i),\mu_c,\Sigma_c,\sigma^2_c)=\alpha_c,$
and from (\ref{eq:meanhills}) we obtain,
$$PG_{61}((\phi_i),\mu_c,\Sigma_c,\sigma^2_c)=
(3/4)\alpha_c+(1/4)
(\ell\Sigma_c^{-1}+C^{-1})^{-1}
\Big[\Sigma_c^{-1}\Big(\sum_i\phi_i\Big)+C^{-1}\eta\Big].$$

\begin{figure}[ht!]
\begin{center}
\includegraphics[width=3.6in]{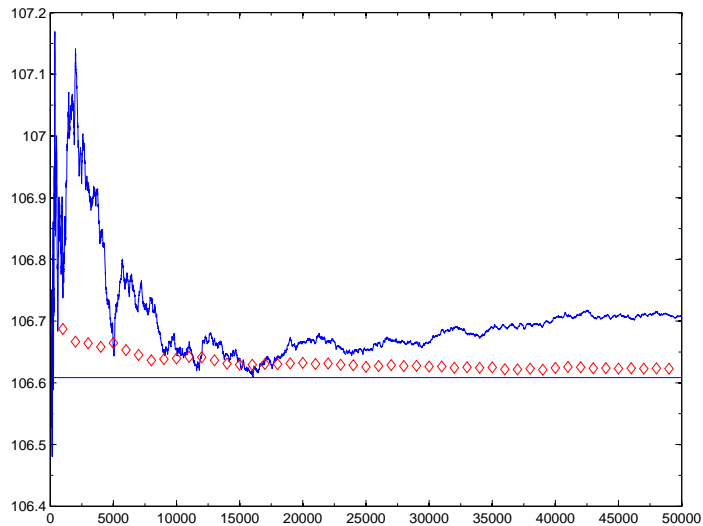}
\caption{The sequence of the standard ergodic averages
is shown as a solid blue line
and the modified estimates
$\mu_{n,\K}(F)$
as red diamonds.
For visual clarity,
the values $\mu_{n,\K}(F)$
are plotted only every 1000 simulation
steps. The ``true'' value of the posterior
mean of $\alpha_c$, shown as a horizontal line,
was computed after $n=10^7$ Gibbs steps and
taken to be $\approx$ 106.6084.}
\label{f:VRF_hn}
\end{center}
\end{figure}

Figure~\ref{f:VRF_hn} shows a typical realization
of the sequence of estimates
obtained by the standard estimators
$\mu_n(F)$ and by the modified estimators
$\mu_{n,\K}(F)$,
for $n=50000$ simulation steps.
The variance of $\mu_{n,\K}(F)$
was found to be 
approximately {\bf 30 times} smaller than 
that of $\mu_n(F)$.
The second row of Table~\ref{tab:rats} shows
the estimated variance reduction factors
obtained at various stages of the MCMC simulation,
based on $T=100$ repetitions of the same experiment,
performed as in Example~1.

The initial values of the sampler were chosen as follows.
For the $(\phi_i)$ we used the ordinary least squares estimates
obtained from $\ell=30$ independent regressions; their sample mean 
and covariance matrix provided starting values for $\mu_c$ and 
$\Sigma_c$, respectively, and a pooled variance estimate 
of the individual regression errors provided the initial 
value of $\sigma_c^2$. The observed data $(Y_{ij})$ and 
known parameter values
for $\eta, C,\nu_0,\rho,R$ and $\tau_0$,
are as in \citet{hills}.

\begin{table}[ht!]
  \begin{center}
    \begin{tabular}{|c||c|c|c|c|c|c|}
      \hline
      \multicolumn{7}{|c|}{\bf Variance reduction factors} \\
      \hline
      & \multicolumn{6}{|c|}{\em Simulation steps}\\
        \hline
{\em Parameter}& $n=1000$& $n=10000$& $n=20000$& $n=50000$& $n=100000$&
							$n=200000$\\
     \hline
$(\phi_i)$     & 1.59-3.58 & 9.12-31.02 & 11.73-61.08 & 10.04-81.36 & 
	12.44-85.99 & 9.38-109.2\\
     \hline
$\alpha_c$     & 2.99 	   & 15.49 	& 32.28	      & 31.14       & 
	28.82	    & 36.48\\
     \hline
$\beta_c$      & 3.05 	   & 19.96 	& 34.05       & 39.22 	    & 
	32.33       & 36.04\\
     \hline
$\Sigma_c$     & 1.15-1.38 & 4.92-5.74  & 5.36-7.60   & 3.88-5.12   & 
	4.91-5.34   & 3.65-6.50\\
     \hline
$\sigma_c^2$   & 2.01 	   & 5.06 	& 5.23 	      & 5.17        & 
	4.75        & 5.79\\
     \hline
     \end{tabular}
  \end{center}
     \caption{Estimated factors by which the
    variance of $\mu_n(F_j)$
    is larger than the 
    variance of
    $\mu_{n,\K}(F)$,
    after
    $n=1000,10000,20000,50000,100000$ and $200000$ 
	simulation
    steps. A different function $F_j$ is defined 
	for each of the $k=66$ scalar parameters in the vector
	$((\phi_i),\mu_c,\Sigma_c,\sigma_c^2)$,
	and the same vector of control variates
	is used for all of them, as specified
	by the basic methodology of Section~\ref{s:basic}.
	In the first row, instead of individual variance
	reduction factors, we state the range of variance
	reduction factors obtained on the 60 individual
	parameters ($\phi_i$), and similarly for the 
	three parameters of $\Sigma_c$ in the fourth row.}
     \label{tab:rats}
\end{table}

More generally, in such a study we would be interested in
the posterior mean of all of the $k=66$ model parameters.
The same experiment as above was performed simultaneously
for all the parameters. Specifically, 66 different functions
$F_j$, $j=1,2,\ldots,66$ were defined, one for each scalar
component of the parameter vector 
$((\phi_i),\mu_c,\Sigma_c,\sigma_c^2)$,
and the modified estimators $\mu_{n,\K}(F_j)$ were used
for each parameter, with respect to the same collection
of control variates as before. The resulting variance
reduction factors (again estimated from $T=100$ repetitions)
are shown in Table~\ref{tab:rats}.


\paragraph{Example 3. Metropolis-within-Gibbs sampling 
from a heavy-tailed posterior. }
We consider an inference problem motivated by
a simplified version of an example
in \citet{roberts-rosenthal:09}.
Suppose $N$ i.i.d.\ observations
$y=(y_1,y_2,\ldots,y_N)$ are drawn from a $N(\phi  ,V)$
distribution, and place independent priors
$\phi  \sim\mbox{Cauchy}(0,1)$ and $V\sim\mbox{IG}(1,1)$,
on the parameters $\phi  ,V$, respectively.
The induced full conditionals of the posterior are
easily seen to satisfy,
\ben
\pi(\phi  |V,y)
&\propto&
    \Big(\frac{1}{1+\phi  ^2}\Big)\exp\Big\{
    -\frac{1}{2V}\sum_i(\phi  -y_i)^2
    \Big\},\\
\mbox{and}\;\;\;\;
\pi(V|\phi  ,y)
&\sim&
    \mbox{IG}
    \Big(1+\frac{N}{2},1+\frac{1}{2}\sum_i(\phi  -y_i)^2\Big).
\een 
Since the distribution $\pi(\phi  |V,y)$ is not of standard
form, direct Gibbs sampling is not possible. Instead, we use a
random-scan Metropolis-within-Gibbs sampler,
cf.\ \citet{muller:93}, \citet{tierney:94}, and either update $V$
from its conditional (Gibbs step), or update $\phi  $ in a random
walk-Metropolis step with a $\phi  '\sim N(\phi  ,1)$ proposal, each
case chosen with probability $1/2$.
Because both the Cauchy and the inverse gamma distributions are 
heavy-tailed, we naturally expect that the MCMC samples will be highly
variable. Indeed, this was found to be the case in the simulation
example considered, where the above algorithm is applied to a
vector $y$ of $N=100$ i.i.d.\ $N(2,4)$ observations, and with
initial values $\phi_0=0$ and $V_0=1$. 
As a result of this
variability, the standard empirical averages of the values of the
two parameters also converge very slowly. Since $V$ is the more
variable of the two, we let $F(\phi  ,V)=V$ and consider the problem
of estimating its posterior mean. 

We compare the performance 
of the standard empirical averages 
$\mu_n(F)$ with that of the 
modified estimators $\mu_{n,\K}(F)$.
As dictated by the basic methodology,
we define $G_1(\phi,V)=\phi$ and $G_2(\phi,V)=V$, 
but we note that the one-step expectation 
$PG_1(\phi,V)$ cannot be obtained analytically
because of the presence of the Metropolis
step. Therefore, we use a single control 
variate $U=G-PG$ defined in terms of the 
basis function $G(\phi,V)=V$. 

Figure~\ref{f:example6b} shows a typical realization of the results
of the two estimators, for $n=10000$ simulation steps. The
corresponding variance reduction factors, estimated from $T=100$
repetitions of the same experiment, are 
{\bf 7.89}, {\bf 7.48}, {\bf 10.46} and {\bf 8.54},
after $n=10000$, $50000$, $100000$ and $200000$
MCMC steps, respectively.
\begin{figure}[ht!]
\centerline{\includegraphics[width=4.4in]{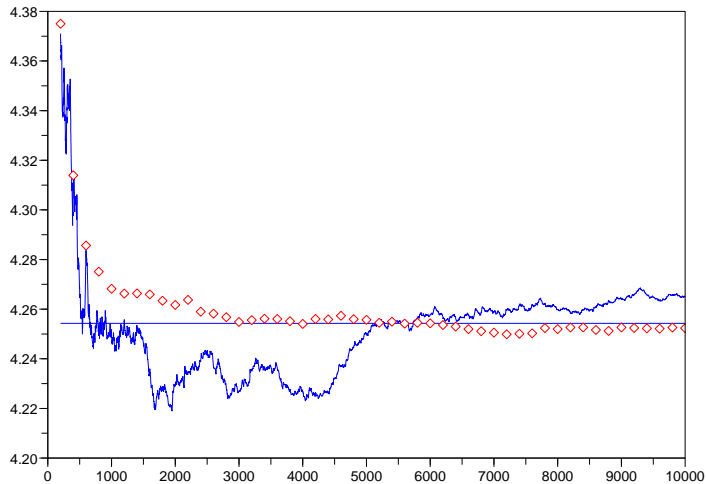}}
\caption{The sequence of the standard ergodic averages
is shown as a solid blue line,
and the modified estimates
$\mu_{n,\K}(F)$ as red diamonds.
For visual clarity,
the values $\mu_{n,\K}(F)$
are plotted only every 200 simulation
steps.  The ``true'' posterior mean of $V$,
plotted as a horizontal line,
is estimated to be $\approx 4.254$
after 10 million simulation steps.
}
\flabel{example6b}
\end{figure}


\section{Extensions of the Basic Methodology}
\label{s:generalMeth}

As discussed in the Introduction, unlike in the case
of independent sampling where control variates need 
to be identified separately in each particular 
application, the basic methodology developed here
provides a simple method for the construction 
of a family of control variates that are 
immediately applicable to a wide range of MCMC 
estimation problems. In particular, it applies 
to any Bayesian inference study where samples from 
the posterior are generated by a conjugate
random-scan Gibbs sampler.

In this section we discuss extensions
of the basic methodology along three directions
where, in each case, we indicate how a different
class of basis functions $\{G_j\}$ can either 
make the use of the associated control variates
more effective, or it can allow their applications
to more general classes of MCMC samplers than
those considered so far. Each of these 
directions is illustrated via 
a detailed MCMC example.

\subsection{`Non-conjugate' samplers}
As noted earlier, the main requirements for the applicability
of the basic methodology of Section~\ref{s:basic}
are that the MCMC sampler be
reversible, and that the conditional expectations
$PG_j(x)=E[G_j(X_{n+1})|X_n=x]$ of (at least some of)
the co-ordinate functions
$G_j$ be explicitly computable. 
Example~4 below illustrates the case of a (non-conjugate)
Gibbs sampler used in the analysis of a log-linear model, 
where, although the functions $PG_j$ cannot be computed
in closed form, it is easy to identify a different
collection of basis functions $\{G_j\}$ for which
the values of the $\{PG_j\}$ are readily available.
With find that, using the resulting control variates
$U_j=G_j-PG_j$ with the modified estimator
in (\ref{eq:Imod}) to estimate the posterior 
means of the model parameters, reduces the 
estimation variance by a factor approximately 
between 3.5 and 180.

\paragraph{Example 4. A log-linear model. }
We consider the $2 \times 3 \times 4$ table presented by
\citet{knuiman}, where $491$ subjects were classified according to
hypertension (yes, no), obesity (low, average, high) and alcohol
consumption (0, 1-2, 3-5, or $6+$ drinks per day).  We choose to
estimate the parameters of the log-linear model with three main
effects and no interactions, specified as,
$$ y_i \sim \mbox{Poisson}(\mu_i),~~\log(\mu_i) = z^t_i \beta,
~~i=1,2\ldots,24,$$
where the $y_i$ denote the cell frequencies,
modeled as Poisson variables with corresponding means
$\mu_i$, each $z_i$ is the $i$th row of the
$24 \times 7$ design matrix $z$ based on sum-to-zero constraints,
and $\beta=(\beta_1,\beta_2,\ldots,\beta_7)^t$ is the parameter vector.
[In \citet{D-forster} this model was identified as having
the highest posterior probability among all log-linear interaction
models, under various prior specifications.]

Assuming a flat prior on $\beta$,
standard Bayesian inference via MCMC can
be performed either by a Gibbs sampler that exploits
the log-concavity of full conditional
densities as in \citet{Dell}, or by a multivariate random walk
Metropolis-Hastings sampler, with a maximum-likelihood
estimate of the covariance matrix offering guidance 
as to the form of the proposal density.  
Instead, here we observe that 
the full conditional density of each $\beta_j$
is the same as the distribution of the {\em logarithm}
of a gamma random variable with density,
\be
\mbox{Gamma} \left( \textstyle{\sum_i} y_i z_{ij},
\sum_{i:z_{ij}=1} \exp \Big\{ \sum_{\ell \neq
j} \beta_\ell z_{i\ell}
\Big\} \right).
\label{eq:logG}
\ee
Hence a (reversible)
random-scan Gibbs sampler 
can be implemented by producing samples
according to the distributions in (\ref{eq:logG})
and taking their logarithm to obtain samples 
from the correct full conditional density
of each $\beta_j$.

In order to estimate the posterior mean 
of every component of $\beta$,
we consider all seven estimation problems
simultaneously, and set $F_j(\beta)=\beta_j$ for all $1\leq j\leq 7$.
But, since the mean of the logarithm of a
gamma-distributed random variable is not known
explicitly, we cannot compute the conditional
expectations $PG_j$ for the co-ordinate functions
$G_j(\beta)=\beta_j$ dictated by the basic methodology.
On the other hand, it is clear 
that the mean of $\exp(\beta_j)$ under the full conditional 
density of $\beta_j$ is simply the mean of 
the gamma density in (\ref{eq:logG}).
Hence, defining basis functions $G_j(\beta)=\exp(\beta_j)$
for each $j=1,2,\ldots,7$, we can simply 
compute,
$$PG_j(\beta)=\frac{6}{7}\,\exp(\beta_j) 
+\frac{1}{7} \left(\frac{\sum_i y_i z_{ij} } {
\sum_{i:z_{ij}=1} \exp \left( \sum_{\ell \neq j} \beta_\ell z_{i\ell}
\right)  }\right).
$$

Therefore, we can use
the same $k=7$ control variates
$U_1,U_2,\ldots,U_7$
for each $F_j$, where the
$U_\ell=G_\ell-PG_\ell$ are defined
in terms of the basis functions above,
in conjunction with the modified estimators
$\mu_{n,\K}(F_j)$ as before.
The variance reduction factors obtained
by $\mu_{n,\K}(F_j)$ after $n=100000$
simulation steps range between 
57.16 and 170.34, for different parameters
$\beta_j$. More precisely, averaging over $T=100$
repetitions, the estimated variance reduction factors 
obtained by $\mu_{n,\K}(F_j)$ are in the range,
{\bf 3.55--5.57, 38.2--57.69, 66.20--135.51,
57.16--170.34} and {\bf 85.41--179.11}, 
after $n=1000, 10000,50000,$ $100000$ and $200000$ 
simulation steps, respectively.
Figure~\ref{f:logl} shows an example of the simulated
results for a sequence of the estimates $\mu_n(F_7)$
and $\mu_{n,\K}(F_7)$ for $\beta_7$.
All MCMC chains were started at the
corresponding maximum likelihood estimates.

\begin{figure}[ht!]
\centerline{\includegraphics[width=4.1in]{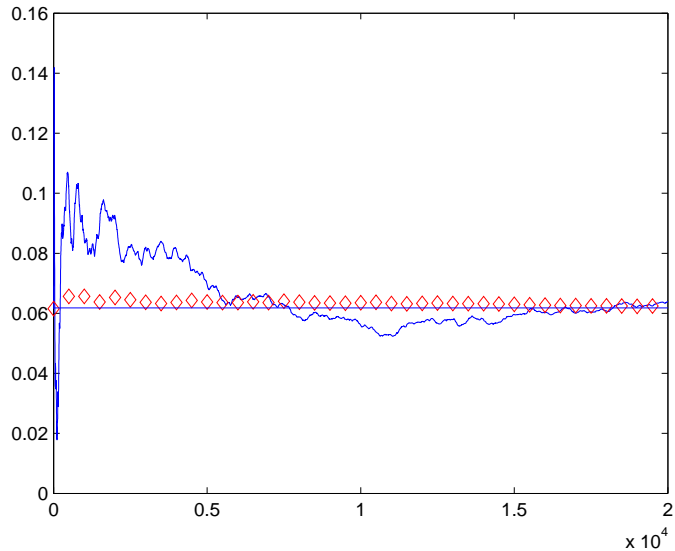}}
\caption{Log-linear model: The sequence of the standard 
estimates $\mu_n(F_7)$ for $\beta_7$ is shown as a solid blue line;
the modified estimates $\mu_{n,\K}(F_7)$, plotted every $500$,
iterations, are plotted as red diamonds. The straight horizontal 
line represents the estimate obtained after $5$ million iterations.}
\flabel{logl}
\end{figure}

%

\subsection{Complex models}
In some estimation problems involving 
more complex models, especially when there is 
reason to expect that the posterior distribution 
is highly non-Gaussian, it may be possible to
identify more effective basis functions $\{G_j\}$ 
than the co-ordinate functions dictated by the
basic methodology, by examining the form
of the associated Poisson equation, as indicated 
by the first rule of thumb of Section~\ref{s:cv2};
that is, by choosing the functions $G_j$ so that 
a linear combination of the form $\sum\theta_jG_j$
approximately satisfies the Poisson equation
for the target function $F$.
One such scenario is illustrated in the following example,
which presents an analysis of a two-component 
Gaussian mixtures model. In order to 
estimate the posterior means of the two location
parameters, a standard random-scan Gibbs 
sampler is employed and the samples are re-ordered
at the post-processing stage. Despite the complex
structure of the resulting chain, the basic 
methodology can still be applied, but it is found
to make little difference in reducing the estimation 
variance. On the other hand, a 
straightforward heuristic analysis of the 
associated Poisson equation indicates that 
the choice of two simple functions $G_1,G_2$
as the basis functions may lead to a fairly
accurate approximation for the solution of 
the associated Poisson equation.
Indeed, the resulting control variates
are found to reduce the estimation variance
by a factor ranging approximately between
15 and 55.

\paragraph{Example 5. Gaussian mixtures. }
Mixtures of densities provide a versatile class of statistical
models that have received a lot of attention from both a theoretical
and a practical perspective.
Mixtures primarily serve as a means of modelling heterogeneity for
classification and discrimination, and as a way of formulating
flexible models for density estimation. Although one of the fist
major success stories in the MCMC community was the Bayesian
implementation of the finite Gaussian mixtures problem
(cf.\ \citet{tanwon87}, \citet{diebolt-robert:94}), 
there are still numerous unresolved
issues in inference for finite mixtures, as discussed, for example,
in the review paper by \citet{Marin05}. These difficulties
emanate primarily from the fact that such models are often ill-posed
or non-identifiable. In terms of Bayesian inference via MCMC, these
issues reflect important problems in prior specifications and label
switching. In particular, improper priors are hard to use and proper
mixing over all posterior modes requires enforcing label-switching
moves through Metropolis steps. Detailed discussions of the dangers
emerging from prior specifications and identifiability constraints
can be found in \citet{Marin05}, \citet{Lee08}, \citet{Jasra05}.

We employ a two-component Gaussian mixtures model.
Starting with $N=500$ data points $y=(y_1,y_2,\ldots,y_N)$
generated from the mixture distribution
$0.7N(0,0.5^2)+0.3N(0.1,3^2)$, and assuming that the
means, variances and mixing proportions are all unknown,
we consider the problem of estimating the
two means. The usual Bayesian formulation introduces
parameters $(\mu_1, \mu_2, \sigma_1^2,\sigma_2^2,Z,p)$, 
as follows.
The data are assumed to be i.i.d.\ from
$pN(\mu_1,\sigma_1^2)+(1-p)N(\mu_2,\sigma_2^2)$,
and we place the following priors:
$p \sim \mbox{Dirichlet}(\delta,\delta)$,
the two means $\mu_1,\mu_2$ are independent
with each $\mu_j \sim N(\xi,\kappa^{-1})$,
and similarly the variances are independent
with each $\sigma_j^{-2} \sim \mbox{Gamma}(\alpha,\beta)$.
We adopt the vague, data-dependent
prior structure of \citet{rich}; 
$\delta$ is set equal to 1, $\xi$ is set equal to 
the empirical mean of the data $y$, 
$\kappa^{-1/2}$ is taken to be equal to the 
data range, $\alpha=2$ and $\beta=0.02\kappa^{-1}$.
Conditional on 
$(\mu_1, \mu_2, \sigma_1^2,\sigma_2^2,p)$,
the latent variables $Z=(Z_1,Z_2,\ldots,Z_N)$
are i.i.d.\ with $\Pr\{Z_i=1\}=1-\Pr\{Z_i=2\}=p,$
and, given the entire parameter vector
$(\mu_1, \mu_2, \sigma_1^2,\sigma_2^2,Z,p)$,
the data $y=(y_i)$ are i.i.d.\
with each $y_i$ having distribution
$N(\mu_j,\sigma_j^2)$ if $Z_i=j$,
for $i=1,2,\ldots,N$, $j=1,2$.

In order to estimate the location parameters
$(\mu_1,\mu_2)$ we sample from
the posterior via a standard random-scan Gibbs sampler,
and we also introduce the {\em a priori} 
restriction that $\mu_1 < \mu_2$. In terms of
the sampling itself, as noted, e.g.,  by \citet{Stephens97}, it is
preferable to first obtain draws from the unconstrained posterior
distribution and then to impose the identifiability constraint at
the post-processing stage. In each iteration, the random-scan Gibbs
sampler selects one of the four parameter blocks $(\mu_1,\mu_2)$,
$(\sigma_1^2,\sigma_2^2)$, $Z$ or $p$, each with probability $1/4$, and
draws a sample from the corresponding full conditional density.
These densities are all of standard form and easy to sample from;
see, for example, \citet{rich}. In particular, the two means are
conditionally independent with,
\be 
\mu_j\sim N\left(
\frac{\sigma_j^{-2}\sum_i(y_i+\kappa\xi)\IND_{\{Z_i=j\}}}
{\sigma_j^{-2}n_j+\kappa}, \frac{1}{\sigma_j^{-2}n_j+\kappa}
\right), \;\;\;\;\mbox{where $n_j=\#\{i:Z_i=j\},\;\;$ $j=1,2$.}
\label{eq:fullCD} 
\ee 
Note that the data $y$ have been generated so that the two
means are very close, resulting in frequent label switching
throughout the MCMC run and in near-identical marginal densities
for $\mu_1$ and $\mu_2$.

We perform a post-processing relabelling of the 
sampled values according to the above restriction
and denote the ordered sampled vector by
$(\mu^o_1, \mu^o_2, \sigma^{o,2}_1,\sigma^{o,2}_2,Z^o,p^o)$.
In order to estimate the posterior mean
of $\mu_1$, let,
$$F(\mu_1, \mu_2, \sigma_1^2,\sigma_2^2,Z,p)
:=\mu_1^o=\min\{\mu_1,\mu_2\}.$$

Since all the one-step conditional expectations 
$PG_j$ of the co-ordinate basis functions $G_j$ 
can be explicitly computed (using the full 
conditional densities of the sampler), 
the basic methodology can be applied directly
in this case. But repeated simulation experiments indicate 
that the gain in variance reduction is generally 
negligible. Specifically, if all the 
co-ordinate functions are used -- one for each parameter 
in the vector $(\mu_1,\mu_2,\sigma_1^2,\sigma_2^2,Z,p)$ --
then $\mu_{n,\K}(F)$ includes a linear combination
of 505 control variates, since there are $N=500$ 
latent variables $Z=(Z_i)$. This introduces a large 
amount of additional variability in $\mu_{n,\K}(F)$,
and its variance is generally larger than that of
$\mu_n(F)$. Moreover, even if the class of
basis functions is restricted to the main five 
parameters $(\mu_1,\mu_2,\sigma_1^2,\sigma_2^2,p)$,
we still find that the use of control variates 
only produces a negligible advantage, if any,
compared to the standard estimates $\mu_n(F)$.

In order to select a potentially more effective
collection $\{G_j\}$ recall that, according
to the first rule of thumb of Section~\ref{s:cv2},
a linear combination of the $G_j$ would ideally
satisfy, at least approximately, the associated
Poisson equation for $F$, namely, 
$P\hat{F}-\hat{F}=-F+\pi(F)=-\mu_1^o+\pi(F)$. 
The first important observation here is that 
the right-hand side of the Poisson equation
is a function of the ordered
sample, whereas all the functions $G_j$ considered
earlier are independent of the ordering of the
two means. Therefore, a natural first candidate
for a basis function is to take
$G_1(\mu_1, \mu_2, \sigma_1^2,\sigma_2^2,Z,p)=
\mu_1^o$; then $PG_1$, the expected value
of $\min\{\mu_1,\mu_2\}$ under (\ref{eq:fullCD}),
is,
\be
&&
PG_1(\mu_1, \mu_2, \sigma_1^2,\sigma_2^2,Z,p)
\;=\;
    \frac{3}{4}G_1(\mu_1, \mu_2, \sigma_1^2,\sigma_2^2,Z,p)
    \nonumber\\
&& \hspace{0.8in}+\;
\frac{\nu_1}{4}\Phi\Big(\frac{\nu_2-\nu_1}{\sqrt{\tau_1^2+\tau_2^2}}\Big)
+\frac{\nu_2}{4}\Phi\Big(\frac{\nu_1-\nu_2}{\sqrt{\tau_1^2+\tau_2^2}}\Big)
-\frac{1}{4}\sqrt{\tau_1^2+\tau_2^2}
\phi\Big(\frac{\nu_2-\nu_1}{\sqrt{\tau_1^2+\tau_2^2}}\Big),
\hspace{0.3in} 
\label{eq:PG1} 
\ee 
where $\nu_j$ and $\tau_j^2$ are
the means and variances of $\mu_j$, respectively, for $j=1,2$, under
the full conditional densities in (\ref{eq:fullCD}); see, for
example, \citet{cain94}. 

If $PG_1-G_1$ were approximately equal to a multiple
of $\mu_1^o$ (up to additive constant terms), we 
could stop here. But (\ref{eq:PG1}) shows that
this difference contains several terms unrelated to 
$\mu_1^0$, with significant additional variability.
So the natural next step is to select $G_2$ so that 
the contribution $PG_2-G_2$ may 
approximately cancel out the last
three terms in the right-hand side of (\ref{eq:PG1}). 
Since the nonlinear terms involving $\phi$ and $\Phi$ 
are hard to handle analytically and are
also bounded, we focus on approximating the
$\sqrt{\tau_1^2+\tau_2^2}$ factor. Note that $\kappa$ 
is typically small compared to $n_1$ and $n_2$, 
so $\tau_1^2$ can be approximated by $\sigma_1^2/(Np)$ 
and $\tau_2^2$ by $\sigma_2^2/(N(1-p))$. 
Finally, since the influence of $\sigma_1^o$ is likely 
to be dominant over that of $\sigma_2^o$ with respect 
to $\mu_1^o$, a straightforward first-order Taylor 
expansion shows that the dominant linear term is
$\sigma_1^o$, suggesting the choice $G_2(\mu_1, \mu_2,
\sigma_1^2,\sigma_2^2,Z,p)=\sigma_1^o$.

To compute $PG_2$, first note that 
the probability that $\mu_1<\mu_2$ under (\ref{eq:fullCD}),
is,
$$p(\mbox{order})
:= \frac{\Phi \big(E( \mu_2|\cdots )- E(\mu_1|\cdots) \big)}
{\sqrt{E(\sigma^2_1|\cdots ) + E(\sigma^2_2|\cdots)}},$$ where all
four expectations above are taken under the corresponding full
conditional densities. Moreover, since the full conditional of each
$\sigma^{-2}_j$ is a gamma density, the expectations of $\sigma_1$,
$\sigma_2$, $\sigma_1^2$, and $\sigma_2^2$, are all available in
closed form. Therefore, $p(\mbox{order})$ can be computed
explicitly, and, \ben &&
    PG_2(\mu_1, \mu_2, \sigma_1^2,\sigma_2^2,Z,p)
    \;=\;\frac{1}{2}G_2(\mu_1, \mu_2, \sigma_1^2,\sigma_2^2,Z,p)\\
&&
    \hspace{0.6in}
    +\; \frac{1}{4} \left[
    \IND_{\{\mu_1<\mu_2\}}
    E(\sigma_1|\cdots)
    +\IND_{\{\mu_1>\mu_2\}}E(\sigma_2|\cdots) \right]
    +
    \frac{1}{4} \Big[ p(\mbox{order}) \sigma_1 +
    (1-p(\mbox{order})) \sigma_2 \Big],
\een
where, again, the expectations are taken under the
corresponding full conditional densities.

With this choice of basis functions $G=(G_1,G_2)^t$,
we define the corresponding control variates
$U=G-PG$, and compare the performance of the
modified estimator $\mu_{n,\K}(F)$ with that
of the standard estimates $\mu_n(F)$.
The resulting variance reduction factors 
(estimated from $T=100$ repetitions of the same
experiment) are {\bf 16.17, 25.36, 38.99, 44.5} 
and  {\bf 36.16}, after $n=1000, 10000,50000,100000$ 
and $200000$ simulation steps,
respectively.
Figure~\ref{f:mixtures} shows 
the results of a
typical simulation run.
The initial values of the sampler were taken
after a 1000-iteration burn-in period, and
the horizontal line in the graph depicting
the ``true'' value of the posterior mean
of $F$ was obtained after 5 million Gibbs iterations.

\begin{figure}[ht!]
\centerline{\includegraphics[width=4.9in]{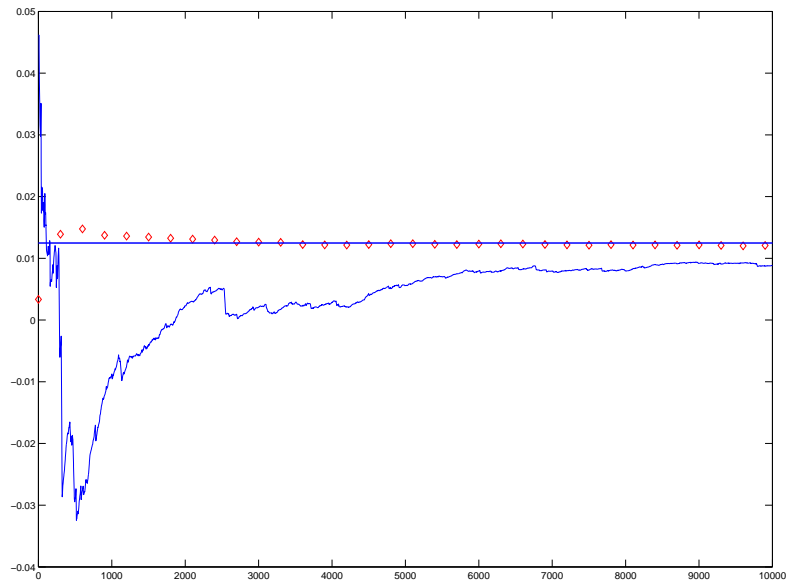}}
\caption{Two-component Gaussian mixture model: The sequence of the standard
ergodic averages for $\mu_1$ is shown as a solid blue line and the
modified estimates $\mu_{n,\K}(F)$,
reported every $300$ iterations,
as red diamonds. } \flabel{mixtures}
\end{figure}

\subsection{General statistics of interest}
In estimation problems where the quantity of 
interest is not the posterior mean of one
of the parameters (so that $F$ itself is not
one of the co-ordinate functions), it is unlikely 
that the co-ordinate functions $\{G_j\}$ would
still lead to effective control variates.
Instead, we argue that it is more appropriate,
in accordance to the second rule of thumb 
of Section~\ref{s:cv2},
to select basis functions $G_j$ so that the 
resulting control variates $U_j:=G_j-PG_j$ 
will be highly correlated with~$F$.
A particular such instance is illustrated 
in Example~6 below, where we examine a
Bayesian model-selection problem arising 
from a two-threshold autoregressive
model after all other parameters have 
been integrated out. [Such applications 
have recently found very intense interest, 
especially in the context of genetics;
see, e.g., \citet{botric}, where the model 
space is endowed with a multimodal discrete
density.]  Here, the  model space is sampled 
by a discrete random-walk Metropolis-Hastings 
algorithm, and the quantity of interest is the 
posterior probability of a particular model.
Defining three basis functions $\{G_j\}$ that
are naturally expected to 
be highly correlated with $F$, 
we find that the variance
of the modified estimators in (\ref{eq:Imod})
is at least 30 times smaller than that of
the standard ergodic averages.

\paragraph{Example 6. A two-threshold autoregressive model. }
We revisit the monthly U.S.\ 3-month treasury bill rates,
from January $1962$ until December $1999$,
previously analyzed by \citet{ddh} using flexible
volatility threshold models.  The time series has $N=456$ points
and is denoted by $r=(r_t)=(r_t\;;\;t=1,2,\ldots,N)$.
Here the data are modeled
in terms of a self-exciting threshold autoregressive
model, with two regimes; it is one of the models proposed
by
\citet{Pfann}, and it is defined as,
\begin{equation}
\Delta r_t = \left\{
\begin{array}{ll}
\alpha_{10} + \alpha_{11} r_{t-1} & r_{t-1} < c_1 \\
\alpha_{20} + \alpha_{21} r_{t-1} & r_{t-1} \geq c_1
\end{array} \right\} + \left\{
\begin{array}{ll}
\sigma_1 \epsilon_t & r_{t-1} < c_2 \\
\sigma_2 \epsilon_t & r_{t-1} \geq c_2
\end{array} \right\},
\label{setar}
\end{equation}
where $\Delta r_t = r_t - r_{t-1}$, and
the parameters $c_1$, $c_2$ are the thresholds
where mean or volatility regime shifts occur.
Instead, we re-write the model as,
\begin{equation}
\Delta r_t =
\left\{
\begin{array}{ll}
\alpha_{10} + \alpha_{11} r_{t-1} & r_{t-1} < c_1 \\
\alpha_{20} + \alpha_{21} r_{t-1} & r_{t-1} \geq c_1
\end{array} \right\} + \left\{
\begin{array}{ll}
\sigma \epsilon_t & r_{t-1} < c_2 \\
\sigma(1+\gamma)^{1/2} \epsilon_t & r_{t-1} \geq c_2
\end{array} \right\},
\label{setar2}
\end{equation}
where $\gamma \geq -1$ characterizes the jump in $\sigma^2$ between
the two volatility regimes. Whereas \citet{Pfann} use a Gibbs
sampler to estimate the parameters of the model in (\ref{setar}), we
exploit the parameterization (\ref{setar2}) as follows.  
Independent improper conjugate priors are adopted 
for the variance, $\pi(\sigma^2) \propto \sigma^{-2}$, 
and for the regression coefficients, $\pi(\alpha_{ij}) \propto 1$,
and the prior for each of $c_1,c_2$ is taken to 
be a discrete uniform distribution over the distinct 
values of $\{r_t\}$, except for the two smallest and 
largest values so that identifiability is obtained; 
the prior for $\gamma$ is chosen to be an exponential 
density with mean one, shifted to $-1$.

The goal is to estimate the posterior
probability $\pi(c_1^1,c_2^1|r)$
of the most likely model, that is, 
of the model corresponding to the pair 
of thresholds $(c_1^1,c_2^1)$ maximizing
$\pi(c_1,c_2|r)$.
In the above formulation, (\ref{setar2}) 
can be written equivalently as,
$
R = X \alpha + \epsilon,
$
where $R = (\Delta r_2, \ldots, \Delta r_T)^t$, $\alpha =
(\alpha_{10}, \alpha_{11}, \alpha_{20}, \alpha_{21})$, $\epsilon$
is a zero-mean Gaussian vector with covariance matrix
$\Sigma,$ and $X$ is the design matrix with row $t$
given by (1 $r_{t-1}$ 0 0), if $r_{t-1} \leq c_1$, and by (0 0 1
$r_{t-1}$) otherwise. The covariance matrix of the errors, $\Sigma$,
is diagonal with $\Sigma_{tt}=\sigma^2$ if
$r_{t-1} \leq c_2$, and $\Sigma_{tt}=(1+\gamma)\sigma^2$, otherwise.
Integrating out the parameters $\alpha$ and $\sigma$,
the marginal likelihood
of the data $r$ with known $c_1$, $c_2$ and $\gamma$
becomes,
$$
p(r | \gamma,c_1,c_2) \propto \exp\Big\{ -\frac{1}{2} \Big[ |\Sigma|
+ \log |X^t \Sigma^{-1} X| + N \log (R^t \Sigma^{-1} R -
\hat{\alpha}^t X^t \Sigma^{-1} X \hat{\alpha}) \Big]\Big\}, $$ where
$\hat{\alpha}=(X^tX)^{-1} X^t R$ is the least-squares estimate of
$\alpha$; see, for example, \citet{OHFOR}. 
After performing
a further one-dimensional numerical integration 
over $\gamma$ by numerical
quadrature, the marginal posterior distribution of
$(c_1,c_2)$ can be written explicitly 
as $\pi(c_1,c_2|r) \propto p(r|c_1,c_2)$.
Therefore, sampling from the posterior of the thresholds
$(c_1,c_2)$ can be performed by a discrete Metropolis-Hastings 
algorithm on $(c_1,c_2)$, where the thresholds $c_1,c_2$ 
take values on the lattice of all the observed values of 
the rates $(r_t)$ except the two farthermost at each end. 
This way, we replace the
$8$-dimensional Gibbs sampler of \citet{Pfann} for~(\ref{setar}), 
by a five-dimensional analytical integration over
$\alpha$ and $\sigma$, a numerical integration over $\gamma$, 
and a Metropolis-Hastings sampler over $(c_1,c_2)$.

Note that this algorithm is computationally less
expensive and also more reliable, since Gibbs
sampling across a discrete and continuous product space 
may encounter ``sticky patches'' in the parameter space.
The discrete Metropolis-Hastings sampler used here
is based on symmetric random walk steps, with vertical 
or horizontal increments of size up to ten, over the 
lattice of all possible values. 
In other words,
the proposed pair $(c'_1,c'_2)$ given the current values $(c_1,c_2)$
is one of the forty neighboring pairs $(c'_1,c'_2)$ of $(c_1,c_2)$,
where two pairs are neighbors if they differ in exactly one
co-ordinate, and by a distance of at most ten locations.
[Clearly, here we do not touch upon the finer issues 
of efficient model searching, as these would
possibly require more sophisticated MCMC algorithms.]

After a preliminary, exploratory simulation stage, the
three {\em a posteriori} most likely pairs of thresholds 
were identified as,
$(c_1^1,c_2^1)=(13.63,\, 2.72)$, $(c_1^2,c_2^2)=(13.89,\, 2.72)$,
$(c_1^3,c_2^3)=(13.78,\, 2.72)$. To estimate the actual
posterior probability of the most likely model,
$\pi(c_1^1,c_2^1|r)$, we define 
$F(c_1,c_2)=\IND_{\{(c_1,c_2)=(c_1^1,c_2^1)\}}$.
For the construction of control variates, as dictated
by the second rule of thumb of Section~\ref{s:cv2},
we select three basis functions $\{G_j\}$ that would
lead to functions $U_j=G_j-PG_j$ which are highly
correlated with $F$. To that end, 
we first make what is perhaps the most obvious choice,
taking $G_1=F$. Then, we define two more basis functions,
according to the same reasoning,
by taking,
$G_j(c_1,c_2)=\IND_{\{(c_1,c_2)=(c_1^j,c_2^j)\}}$, for $j=2,3$.

Observe that, here, since the number of all possible random-walk
moves is quite limited, the one-step expectations $PG_j(x)$ can 
be written out explicitly and their required values at specific
points $x=(c_1,c_2)$ can easily be computed:
Indeed, writing $(c_1,c_2)\sim(c_1',c_2')$ when $(c_1,c_2)$ and
$(c_1',c_2')$ are neighboring pairs, $PG_j$ can be expressed,
for $j=1,2,3$,
$$PG_j(c_1,c_2)=
\begin{cases}
1-\frac{1}{40}\sum_{(c'_1,c'_2)\sim(c_1,c_2)}
    \min\Big\{1, \frac{p(r|c'_1,c'_2)}{p(r|c_1,c_2)}\Big\},
    & \text{if $(c_1,c_2)=(c_1^j,c_2^j)$;}\\
\frac{1}{40}
    \min\Big\{1, \frac{p(r|c^j_1,c^j_2)}{p(r|c_1,c_2)}\Big\},
    & \text{if $(c_1,c_2)\sim (c_1^j,c_2^j)$;}\\
0,
    & \text{otherwise.}\\
\end{cases}
$$

The resulting variance reduction factors 
obtained by $\mu_{n,\K}(F)$,
estimated from $T=100$ repetitions,
are {\bf 125.16, 32.83, 36.76, 30.90} and  {\bf 30.11},
after $n=10000, 20000,50000,100000$ and $200000$ 
simulation steps, respectively.
Figure \ref{f:finance} shows a typical simulation run.
All MCMC chains were started at $(c_1^1, c_2^1)$.

\begin{figure}[ht!]
\centerline{\includegraphics[width=4.1in]{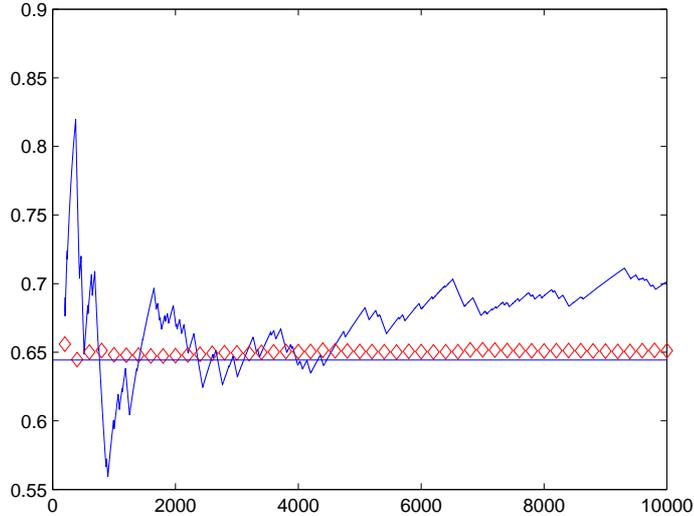}}
\caption{Two-threshold autoregressive model:
The sequence of the standard ergodic averages $\mu_n(F)$ is
shown as a solid blue line and the modified estimates
$\mu_{n,\K}(F)$, reported every $500$ iterations, as red
diamonds. The straight horizontal line represents the posterior
model probability obtained after $50$ million iterations.}
\flabel{finance}
\end{figure}

\newpage

\section{Further Methodological Issues}
\label{s:issues}

Here we examine a number of further methodological 
extensions to the results presented so far,
and we discuss a few different, related approaches.

\subsection{Alternative consistent estimators for $\theta^*$}
\label{s:Gamma}
Recall that the estimator $\hat{\theta}_{n,\K}$ for 
the optimal coefficient vector $\theta^*=(\theta_1^*,\theta_2^*,
\ldots,\theta_k^*)^t$ defined in Section~\ref{s:thetastar} 
was motivated by the new representation for $\theta^*$ 
derived in equation~(\ref{eq:theta_rev_k_2}) of Proposition~2. 
But we also derived an alternative expression 
for $\theta^*$ in equation~(\ref{eq:theta_rev_k}),
as
$\theta^*=\Gamma(G)^{-1}
\pi\big((F-\pi(F))(G+PG)\big),$
where 
$\Gamma(G)_{ij}=\pi(G_iG_j-(PG_i)(PG_j))$,
$1\leq i,j\leq k$.
This suggests that $\theta^*$ can alternatively
be estimated via,
\ben
\hat{\theta}_{n,\Gamma}
:=
    \Gamma_n(G)^{-1}
    [\mu_n(F(G+PG))-
    \mu_n(F)\mu_n(G+PG)],
\een
where the empirical $k\times k$
matrix $\Gamma_n(G)$ is
defined by,
$(\Gamma_n(G))_{ij}
=
\mu_n(G_iG_j-(PG_i)(PG_j))$, 
$1\leq i,j\leq k.$ 
Then $\hat{\theta}_{n,\Gamma}$ can in
turn be used in conjunction with the
vector of control variates $U=G-PG$
to estimate $\pi(F)$ via,
\be
\mu_{n,\Gamma}(F)
:=
    \mu_n(F_{\hat{\theta}_{n,\Gamma}})
    \;=\;\mu_n(F)-\langle\hat{\theta}_{n,\Gamma},\mu_n(U)\rangle
    \;=\;\frac{1}{n}\sum_{i=0}^{n-1}
	\Big[
	F(X_i)-\langle\hat{\theta}_{n,\Gamma},U\rangle
	\Big] .
    \label{eq:multi-est2-2}
\ee

In theory, the estimators 
$\hat{\theta}_{n,\Gamma}$ and $\mu_{n,\Gamma}(F)$ 
enjoy the exact same asymptotic
consistency and optimality properties as their
earlier counterparts, $\hat{\theta}_{n,\K}$ and 
$\mu_{n,\K}(F)$, respectively; these are established
in Section~\ref{s:theory}.
Also, the overall empirical performance
of $\hat{\theta}_{n,\Gamma}$ and $\mu_{n,\Gamma}(F)$ 
was found in practice to be very similar
to that of $\hat{\theta}_{n,\K}$ and $\mu_{n,\K}(F)$.
This was observed in numerous simulation experiments 
we conducted, some of which are reported in the 
unpublished notes of \citet{dellaportas-K:arxiv}.
A small difference between the two estimators was observed 
in experiments where the initial values of the 
sampler were quite far from the bulk of the mass 
of the target distribution $\pi$. There
$\hat{\theta}_{n,\Gamma}$ sometimes appeared 
to converge faster than $\hat{\theta}_{n,\K}$,
and the corresponding estimator
$\mu_{n,\Gamma}(F)$ often gave somewhat better 
results than $\mu_{n,\K}(F)$. The reason for
this discrepancy is the existence of a
time-lag in the definition of
$\hat{\theta}_{n,\K}$: When the initial simulation 
phase produces samples that approach the area near 
the mode of $\pi$ approximately monotonically,
the entries of the matrix $\K_n(G)$ accumulate a systematic 
one-sided error and consequently $\K_n(G)$ takes longer 
to converge than $\Gamma_n(G)$. But this is a transient phenomenon that
can be easily eliminated by including a burn-in 
phase in the simulation.

One the other hand, we 
systematically observed
that the estimator $\hat{\theta}_{n,\K}$
was more stable than $\hat{\theta}_{n,\Gamma}$, 
especially so in the more complex MCMC scenarios 
involving a larger number of control variates. 
This difference was particularly pronounced 
in cases where one
or more of the entries on the diagonal of 
$\Gamma(G)=\K(G)$ were near zero. There,
because of the inevitable fluctuations
in the estimation process,
the values of some of the entries 
of $\hat{\theta}_{n,\Gamma}$
fluctuated wildly between large negative
and large positive values, whereas the 
corresponding entries of $\hat{\theta}_{n,\K}$ 
were much more reliable since, by definition, 
$\K(G)$ is positive semidefinite.

In conclusion, we found that, among the two
estimators, $\mu_{n,\K}(F)$ was consistently 
the more reliable, preferable choice.

We also briefly mention that a different method 
for consistently estimating $\theta^*$ was recently 
developed in \citet{meyn:book}, based on the 
``temporal difference learning'' algorithm. 
Although this method also applies to non-reversible 
chains, it is computationally significantly more 
expensive than the estimates 
$\hat{\theta}_{n,\K}$ and $\hat{\theta}_{n,\Gamma}$,
and its applicability is restricted to 
discrete-state space chains (or,
more generally, to chains 
containing an atom).
It may be possible to extend this idea to 
more general classes of chains by a simulated
construction analogous to Nummelin's 
``split chain'' technique, cf.\
\citet{nummelin:book}, but we have not
pursued this direction further.

\subsection{Estimating $\theta^*$ via batch-means}
\label{s:batch}
As noted in Section~\ref{s:cv2},
the main difficulty in estimating the 
optimal coefficient vector $\theta^*$ 
in~(\ref{eq:thetaStar-k}) was that it
involves the solution $\hat{F}$ to the Poisson 
equation. Various authors have observed 
in the past (see the references below) that 
one possible way to overcome this 
problem is to note that $\theta^*$ (like $\hat{F}$
itself) can alternatively be 
written in terms of an infinite series. 
Restricting attention, for the sake
of simplicity, to the case of a single control 
variate $U=G-PG$ based on a single function 
$G:\state\to\RL$, from equation~(\ref{eq:thetaRatio})
we have,
\be
\theta^*
\;=\;\frac
{1}{\sigma_U^2}
\sum_{j=-\infty}^\infty E_\pi[F(X_0)U(X_j)]
\;=\;\frac
{1}{\pi(G^2-(PG)^2)}
\sum_{j=-\infty}^\infty E_\pi[F(X_0)U(X_j)].
\label{eq:series}
\ee

This suggests the following simple strategy: Truncate
the series in~(\ref{eq:series}) to a finite sum,
from $j=-M$ to $j=M$, say, and estimate an approximation
to $\theta^*$ via,
\be
\tilde{\theta}_{n,M}=\frac{1}{\mu_n(G^2-(PG)^2)}\sum_{j=-M}^M
\frac{1}{n-2M}\sum_{i=M+1}^{n-M}F(X_i)U(X_{i+j}).
\label{eq:thetabatch}
\ee
Then, $\tilde{\theta}_{n,M}$ converges a.s.\ to,
\be
\tilde{\theta}_M:=
\frac{1}{\sigma_U^2}
\sum_{j=-M}^M E_\pi[F(X_0)U(X_j)],
\label{eq:thetaM}
\ee
as $n\to\infty$ (see Corollary~2 in Section~\ref{s:theory}), 
and one would hope that
$\tilde{\theta}_M\approx\theta^*$
for ``large enough'' $M$.
Using the estimated coefficient $\tilde{\theta}_{n,M}$
in conjunction with the control variate $U=G-PG$, 
$\pi(F)$ can then be estimated by the corresponding
modified averages,
\be
\tilde{\mu}_{n,M}(F):=\mu_n(F_{\tilde{\theta}_{n,M}})
=\mu_n(F)-\tilde{\theta}_{n,M}\mu_n(U).
\label{eq:mubatch}
\ee

This methodology has been used extensively 
in the literature, including, among others,
by \citet{adra}, 
\citet{Mira_tecrep},
\citet{stein-diaconis-et-al},
\citet{meyn:book} and
\citet{Hammer08}.
Our main point
here is to show that it is strictly
suboptimal, and in certain cases severely so.
To that end, we next give a precise expression
for the amount by which the asymptotic variance
of the batch-means estimators
$\tilde{\mu}_{n,M}(F)$ is {\em larger} than
the (theoretically minimal) variance $\sigma_{\theta^*}^2$
of $\mu_{n,\K}(F)$. Proposition~3 is proved at the
end of this section.

\medskip

\noindent
{\bf Proposition 3. } {\em The sequences
of estimators
$\{\tilde{\mu}_{n,M}(F)\}$ and 
$\{\mu_{n,\K}(F)\}$ are both 
asymptotically normal: As $n\to\infty$,
\ben
\sqrt{n}[\tilde{\mu}_{n,M}(F)-\pi(F)]
&\tod&
	N(0,\tau_M^2)\\
\sqrt{n}[\mu_{n,\K}(F)-\pi(F)]
&\tod&
	N(0,\sigma_{\theta^*}^2),
\een
where `$\tod$' denotes convergence in
distribution. Moreover, the difference
between the variance
of the batch-means estimators
$\tilde{\mu}_{n,M}(F)$ and that
of the modified estimators $\mu_{n,\K}(F)$ is,
\be
\tau_M^2-\sigma_{\theta^*}^2\;=\;\frac{1}{\sigma_U^2}
\Big[\sum_{|j|\geq M+1}E_\pi[F(X_0)U(X_j)]\Big]^2
\geq 0.
\label{eq:loss}
\ee
}

It is evident from~(\ref{eq:loss}) that 
the variance $\tau_M^2$ of the batch-means 
estimators $\tilde{\mu}_{n,M}(F)$ will often 
be significantly larger than the minimal 
variance $\sigma_{\theta^*}^2$ achieved by $\mu_{n,\K}(F)$.
Especially so if either:
($i$)~The MCMC samples are highly correlated 
(as is often the case with samplers that tend
to make small, local moves), so that the terms 
of the series $\sum_j E_\pi[F(X_0)U(X_j)]$ decay 
slowly with $|j|$; or, ($ii$)~$|F|$ tends to take
on large values; e.g., note that the difference 
in~(\ref{eq:loss}) can be made arbitrarily large 
by multiplying $F$ by a big constant.
But these are exactly the two most common situations
that call for the use of a variance reduction 
technique such as control variates.

Indeed, in numerous simulation experiments 
(some simple cases of which are reported in the 
unpublished notes of \citet{dellaportas-K:arxiv}) 
we observed that,
compared to $\mu_{n,\K}(F)$, 
the batch-means estimators 
$\tilde{\mu}_{n,M}(F)$ require significantly more 
computation (especially for large $M$) and they 
are typically much less effective.
Also, we are unaware of any 
reasonably justified (non-{\em ad hoc}) guidelines 
for the choice of the parameter $M$, which is 
critical for any potentially useful
application of $\tilde{\mu}_{n,M}(F)$.

Using the obvious extension of the above construction
to the case when more than a single control variate
is used, we re-examined the three examples of the
basic methodology that were presented in Section~\ref{s:simple}.
Tables~\ref{tab:batch1},~\ref{tab:batch2} and~\ref{tab:batch3} below
show the results obtained by the batch-means estimators
$\tilde{\mu}_{n,M}(F)$ for various choices of $M$,
together with the earlier results obtained by $\mu_{n,\K}(F)$.
In each case, the sampling
algorithm and all relevant parameters are as in the
corresponding example in Section~\ref{s:simple}. 
In the interest of space, for Example~2 
(Table~\ref{tab:batch2}) we only display results 
for the problem of estimating what is probably 
the statistically most significant parameter 
in this study, namely,
the mean slope $\beta_c$, which corresponds to
taking $F((\phi_i),\mu_c,\Sigma_c,\sigma^2_c)=\beta_c$.

\begin{table}[ht!]
  \begin{center}
    \begin{tabular}{|c||c|c|c|c|}
      \hline
      \multicolumn{5}{|c|}{\bf Variance reduction factors for Example 1} \\
      \hline
      & \multicolumn{4}{|c|}{\em Simulation steps}\\
        \hline
{\em Estimator}& 
	$n=1000$& $n=10000$& $n=50000$& $n=500000$\\
     \hline
$\tilde{\mu}_{n,M}(F),\;M=0$     
 		    & 1.01 & 1.01 & 1.01 & 1.01 \\
     \hline
$\tilde{\mu}_{n,M}(F),\;M=1$     
		    & 1.02 & 1.02 & 1.01 & 1.02 \\
     \hline
$\tilde{\mu}_{n,M}(F),\;M=5$     
		    & 1.06 & 1.06 & 1.06 & 1.06 \\
     \hline
$\tilde{\mu}_{n,M}(F),\;M=10$     
		    & 1.12 & 1.11 & 1.11 & 1.11 \\
     \hline
$\tilde{\mu}_{n,M}(F),\;M=20$     
		    & 1.26 & 1.23 & 1.23 & 1.23 \\
     \hline
$\tilde{\mu}_{n,M}(F),\;M=100$     
		    & 1.64 & 2.78 & 2.78 & 2.74 \\
     \hline
$\tilde{\mu}_{n,M}(F),\;M=200$     
		    & 1.88 & 8.77 & 7.57 & 7.44 \\
     \hline
$\mu_{n,\K}(F)$     
	& {\bf 4.13} & {\bf 27.91} & {\bf 122.4} 
	& {\bf 1196.6}\\
     \hline
     \end{tabular}
  \end{center}
     \caption{Estimated factors by which the
    variance of $\mu_n(F)$ is larger than the 
    variances of $\tilde{\mu}_{n,M}(F)$ and $\mu_{n,\K}(F)$.
	All estimators are applied to data generated 
	by the random-scan Gibbs sampler
	for a highly correlated bivariate Gaussian density
	with parameters chosen as in Example~1. 
	Results are shown after $n=1000,10000,50000$ and $500000$ 
	simulation steps, when the batch-means parameter 
	$M=0,1,5,10,20,100$ or $200$. The variance reduction
	factors are computed from $T=100$ independent repetitions
	of the same experiment. [See the description in Example~1.]
	}
     \label{tab:batch1}
\end{table}

\begin{table}[ht!]
  \begin{center}
    \begin{tabular}{|c||c|c|c|c|}
      \hline
      \multicolumn{5}{|c|}{\bf Variance reduction factors for Example 2} \\
      \hline
      & \multicolumn{4}{|c|}{\em Simulation steps}\\
        \hline
{\em Estimator}& 
	$n=1000$& $n=10000$& $n=50000$& $n=200000$\\
     \hline
$\tilde{\mu}_{n,M}(F),\;M=0$     
 		    & 1.00      & 1.00 & 1.00 & 1.00 \\
     \hline
$\tilde{\mu}_{n,M}(F),\;M=1$     
		    & 0.94      & 1.00 & 1.00 & 1.00 \\
     \hline
$\tilde{\mu}_{n,M}(F),\;M=5$     
		    & 0.24      & 1.00 & 1.00 & 0.99\\
     \hline
$\tilde{\mu}_{n,M}(F),\;M=10$     
		    & 0.09      & 1.00 & 1.00 & 0.99\\
     \hline
$\tilde{\mu}_{n,M}(F),\;M=20$     
		    & 0.45      & 0.99 & 1.00 & 0.98\\
     \hline
$\tilde{\mu}_{n,M}(F),\;M=100$     
		    & $10^{-4}$ & 0.84 & 0.95 & 0.96\\
     \hline
$\mu_{n,\K}(F)$     
	& {\bf 3.05} & {\bf 19.96} & {\bf 39.22} 
	& {\bf 36.04}\\
     \hline
     \end{tabular}
  \end{center}
     \caption{ 
	Estimated factors by which the variance of 
	$\mu_n(F)$ is larger than the 
    	variances of $\tilde{\mu}_{n,M}(F)$ and $\mu_{n,\K}(F)$.
	All estimators are
	applied to MCMC data sampled from the posterior 
	of the hierarchical
	linear model described in Example~2,
	and the parameter being estimated is $\beta_c$
	so that here
	$F((\phi_i),\mu_c,\Sigma_c,\sigma^2_c)=\beta_c$.
	Results are shown after $n=1000,10000,50000$ and $200000$ 
	simulation steps, with the
	batch-means parameter
	$M=0,1,5,10,20$ or $100$. The variance reduction
	factors are computed from $T=100$ independent repetitions
	of the same experiment.
	}
     \label{tab:batch2}
\end{table}

\begin{table}[ht!]
  \begin{center}
    \begin{tabular}{|c||c|c|c|c|}
      \hline
      \multicolumn{5}{|c|}{\bf Variance reduction factors for Example~3} \\
      \hline
      & \multicolumn{4}{|c|}{\em Simulation steps}\\
        \hline
{\em Estimator}& $n=10000$& $n=50000$& $n=100000$& $n=200000$\\
     \hline
$\tilde{\mu}_{n,M}(F),\;M=0$     
 		    & 1.99 & 1.95 & 1.99 & 1.97\\
     \hline
$\tilde{\mu}_{n,M}(F),\;M=1$     
		    & 3.39 & 4.14 & 3.96 & 4.65\\
     \hline
$\tilde{\mu}_{n,M}(F),\;M=5$     
		    & 3.86 & 5.72 & 7.69 & 5.22 \\
     \hline
$\tilde{\mu}_{n,M}(F),\;M=10$     
		    & 0.44 & 3.41 & 4.21 & 5.99\\
     \hline
$\tilde{\mu}_{n,M}(F),\;M=20$     
		    & 0.15 & 0.90 & 2.21  & 3.18\\
     \hline
$\mu_{n,\K}(F)$     & {\bf 7.89} & {\bf 7.48} & {\bf 10.4} & {\bf 8.54} \\
     \hline
     \end{tabular}
  \end{center}
     \caption{
	Estimated factors by which the variance of 
	$\mu_n(F)$ is larger than the 
    	variances of $\tilde{\mu}_{n,M}(F)$ and $\mu_{n,\K}(F)$.
	All estimators are
	applied to data generated by the Metropolis-within-Gibbs
	sampler of Example~3, simulating 
	a simple heavy-tailed posterior.
	Results are shown after $n=10000,50000,100000$ and $200000$ 
	simulation steps, for the following values 
	of the batch-means parameter 
	$M=0,1,5,10$ and $20$. The variance reduction
	factors are computed from $T=100$ independent repetitions
	of the same experiment.
	}
     \label{tab:batch3}
\end{table}

\newpage

\medskip

\noindent
{\sc Proof of Proposition~3. }
The asymptotic normality statements are established 
in Corollaries~1 and~2 of Section~\ref{s:theory}.
To simplify the notation we decompose the
infinite series in~(\ref{eq:series}) into the
sum $S_M+T_M$, where $S_M$ is the sum of the terms
corresponding to $-M\leq j\leq M$ and $T_M$ is
the double-sided tail series corresponding to the 
same sum over all $|j|\geq M+1$. The variances of the two
estimators are given by,
\ben
\tau_M^2
&=&
	\sigma_F^2+\tilde{\theta}_M^2\sigma_U^2
	-2\tilde{\theta}_M(S_M+T_M)\\
\sigma^2_{\theta^*}
&=&
	\sigma_F^2 -\frac{1}{\sigma_U^2}
	(S_M+T_M)^2,
\een
cf.~(\ref{eq:tauSM}) and~(\ref{eq:thetaseries}), respectively.
Taking the difference between the two and substituting 
the value of $\tilde{\theta}_M=S_M/\sigma_U^2$ 
from~(\ref{eq:thetaM}), gives,
$$
\tau_M^2-\sigma_{\theta^*}^2
=
\frac{1}{\sigma_U^2}
[S_M^2-2S_M(S_M+T_M)+(S_M+T_M)^2]
=
\frac{1}{\sigma_U^2} T_M^2,
$$
as claimed in~(\ref{eq:loss}).
\qed

In view of Corollary~1 in Section~\ref{s:theory},
a simple examination of the above proof shows that 
the result of Proposition~3 also holds with the 
estimators $\mu_{n,\Gamma}(F)$ introduced 
in Section~\ref{s:Gamma} in place of $\mu_{n,\K}(F)$.

\newpage

\section{Theory}
\label{s:theory}

In this section we give precise conditions under which the
asymptotics developed in Sections~\ref{s:cv},~\ref{s:thetastar}
and~\ref{s:batch} are rigorously
justified. The results together with their
detailed assumptions are stated below
and the proofs are contained in the appendix.

First we recall the basic setting from \Section{cv}.
We take $\{X_n\}$ to be a Markov chain with values in
a general measurable space $\state$ equipped with a
$\sigma$-algebra $\clB$. The distribution of $\{X_n\}$
is described by its initial state $X_0=x\in\state$
and its transition kernel, $P(x,dy)$, as in (\ref{eq:kernel}).
The kernel $P$, as well as any of its powers $P^n$,
acts linearly on functions $F:\state\to\RL$ via,
$PF(x)=E[F(X_1)|X_0=x]$.

Our first assumption on the chain $\{X_n\}$ is that
it is
\textit{$\psi$-irreducible} and \textit{aperiodic}.
This means that there is a $\sigma$-finite measure
$\psi$ on $(\state,\clB)$ such that,
for any $A\in\clB$ satisfying $\psi(A)>0$
and any initial condition $x$,
\[
P^n(x,A) > 0,\qquad \hbox{for all $n$ sufficiently large.}
\]
Without loss of generality, $\psi$ is assumed to
be \textit{maximal} in the sense that any other
such $\psi'$ is absolutely continuous with respect
to $\psi$.

Our second, and stronger, assumption, is an essentially minimal
ergodicity condition; cf. \citet{meyn-tweedie:book2}: We assume that
there are functions $V:\state\to [0,\infty)$, $W:\state\to
[1,\infty)$, a ``small'' set $C\in\clB$, and a finite constant $b>0$
such that the Lyapunov drift condition (V3) holds:
$$PV-V\leq - W+b\IND_C.
\eqno{\hbox{(V3)}}
$$
Recall that a set $C\in\clB$ is {\em small} if there exists an integer
$m\geq 1$, a $\delta>0$ and a probability measure $\nu$ on
$(\state,\clB)$ such that,
$$P^m(x,B)\geq\delta\nu(B)\;\;\;\;\mbox{for all }\;
x\in C,\;B\in\clB.$$
Under (V3), we are assured that the chain
is positive recurrent and that it possesses
a unique invariant (probability) measure
$\pi$. Our final assumption on the chain is that the
Lyapunov function $V$ in (V3) satisfies,
$\pi(V^2)<\infty$.

These assumptions are summarized as follows:
$$
\left.
\mbox{\parbox{.73\hsize}{\raggedright
    The chain $\{X_n\}$ is $\psi$-irreducible
    and aperiodic, with unique invariant measure
    $\pi$, and there exist functions
    $V:\state\to [0,\infty)$, $W:\state\to [1,\infty)$,
    a small set $C\in\clB$, and a finite constant
    $b>0$, such that (V3) holds and $\pi(V^2)<\infty$.
    }}
\;\;\right\}
\eqno{\mbox{(A)}}
$$
Although these conditions may seem somewhat involved,
their verification is generally straightforward;
see the texts by \citet{meyn-tweedie:book2} 
and by \citet{robert:book},
as well as the numerous examples developed in
\citet{roberts-tweedie:96}, \citet{hobert-geyer:98},
\citet{jarner-hansen:00}, \citet{fort-et-al:03},
\citet{roberts-rosenthal:04}. It is often possible
to avoid having to verify (V3) directly, by appealing to the
property of geometric ergodicity, which is essentially equivalent to
the requirement that (V3) holds with $W$ being a multiple of the
Lyapunov function $V$. For large classes of MCMC samplers, geometric
ergodicity has been established in the above papers, among others.
Moreover, geometrically ergodic chains, especially in the reversible
case, have many attractive properties, as discussed, for example, by
\citet{roberts-rosenthal:98}.

In the interest of generality, the main
results of this section are stated in terms of the weaker
(and essentially minimal) assumptions in~(A).
Some details on general strategies for their
verification can be found in the references above.

Apart from conditions on the Markov chain $\{X_n\}$,
the asymptotic results stated earlier also require
some assumptions on the function $F:\state\to\RL$
whose mean under $\pi$ is to be estimated,
and on the (possibly vector-valued) function
$G:\state\to\RL^k$ which is used for the
construction of the
control variates $U=G-PG$. These assumptions
are most conveniently stated within the
weighted-$L_\infty$ framework of
\citet{meyn-tweedie:book2}.
Given an arbitrary function
$W:\state\to[1,\infty)$, the weighted-$L_\infty$
space $L_\infty^W$ is the Banach space,
$$L_\infty^W:=\Big\{\mbox{functions}\;F:\state\to \RL\;\;
\mbox{s.t.}\;\;\|F\|_W:=\sup_{x\in\state}
\frac{|F(x)|}{W(x)}<\infty\Big\}.$$
With a slight abuse of notation, we
say that a vector-valued function
$G=(G_1,G_2,\ldots,G_k)^t$ is in $L_\infty^W$
if $G_j\in L_\infty^W$ for each $j$.

\medskip

\noindent
{\bf Theorem 2. }
{\em
Suppose the chain $\{X_n\}$
satisfies conditions {\em (A)}, and let
$\{\theta_n\}$ be any
sequence of random vectors in $\RL^k$
such that $\theta_n$ converge to some
constant $\theta\in\RL^k$
a.s., as $n\to\infty$.
Then:
\begin{enumerate}
\item[{\bf (i)}]
{\sc [Ergodicity] }
The chain is positive Harris recurrent, it  has a unique
invariant (probability) measure $\pi$, and it
converges in distribution to $\pi$,
in that for any $x\in\state$ and $A\in\clB$,
$$P^n(x,A)\to\pi(A),\;\;\;\;\mbox{as }\;n\to\infty.$$
In fact, there
exists a finite constant $B$ such that,
\be
\sum_{n=0}^\infty |P^nF(x)-\pi(F)|\leq B(V(x)+1),
\label{eq:V3bound}
\ee
uniformly over all initial states $x\in\state$
and all function $F$ such that $|F|\leq W$.
\item[{\bf (ii)}]
{\sc [LLN] }
For any $F,G\in L_\infty^W$ and any $\vartheta\in\RL^k$,
write $U=G-PG$ and $F_\vartheta:=F-\langle\vartheta,U\rangle$.
Then
the ergodic averages $\mu_n(F)$,
as well as the
modified averages
$\mu_n(F_{\theta_n})$,
both converge to $\pi(F)$ a.s.,
as $n\to\infty$.
\item[{\bf (iii)}]
{\sc [Poisson Equation] }
If $F\in L_\infty^W$, then there exists
a solution $\hat{F}\in L_\infty^{V+1}$ to
the Poisson equation, $P\hat{F}-\hat{F}=-F+\pi(F)$,
and $\hat{F}$ is unique up to an additive constant.
\item[{\bf (iv)}]
{\sc [CLT for $\mu_n(F)$] }
If $F\in L_\infty^W$ and the variance,
$\sigma_{F}^2:=\pi(\hat{F}^2-(P\hat{F})^2)$ is nonzero,
then the normalized ergodic averages
$\sqrt{n}[\mu_n(F)-\pi(F)]$
converge in distribution to $N(0,\sigma_F^2)$,
as $n\to\infty$.
\item[{\bf (v)}]
{\sc [CLT for $\mu_n(F_{\theta_n})$] }
If $F,G\in L_\infty^W$,
and the variances,
$\sigma_{F_{\theta}}^2:= \pi(\hat{F}_{\theta}^2-(P\hat{F}_{\theta})^2)$
and
$\sigma_{U_j}^2:= \pi(\hat{U}_j^2-(P\hat{U}_j)^2)$,
$j=1,2,\ldots,k$ are all
nonzero, then the normalized modified
averages $\sqrt{n}[\mu_n(F_{\theta_n})-\pi(F)]$
converge in distribution to $N(0,\sigma_{F_{\theta}}^2)$,
as $n\to\infty$.
\end{enumerate}
}

\medskip

Suppose the chain $\{X_n\}$ satisfies conditions
(A) above, and that the functions $F$
and $G=(G_1,G_2,\ldots,G_k)^t$ are
in $L_\infty^W$. Theorem~2 states that
the ergodic averages $\mu_n(F)$ as well
as the modified averages
$\mu_n(F_\theta)$ based on the
vector of control variates
$U=G-PG$ both converge to $\pi(F)$,
and both are asymptotically normal.

Next we examine the choice of
the coefficient vector $\theta=\theta^*$
which minimizes the limiting variance
$\sigma_{F_\theta}^2$
of the modified averages, and the
asymptotic behavior of
the estimators $\hat{\theta}_{n,\Gamma}$
and $\hat{\theta}_{n,\K}$ for $\theta^*$.

As in Section~\ref{s:cv2}, let
$\Gamma(G)$ denote the $k\times k$
matrix with entries,
$\Gamma(G)_{ij}=\pi(G_iG_j-(PG_i)(PG_j))$,
and recall that, according to
Theorem~2, there exists a solution $\hat{F}$
to the Poisson equation for $F$. The simple
computation outlined in Section~\ref{s:cv2}
(and justified in the proof of Theorem~3)
leading to equation (\ref{eq:thetaStar-k})
shows that the variance $\sigma_{F_\theta}^2$
is minimized by the choice,
\ben
\theta^*=
\Gamma(G)^{-1}\pi(\hat{F}G-(P\hat{F})(PG)),
\een
as long as the matrix $\Gamma(G)$ is invertible.
Our next result establishes the a.s.-consistency
of the estimators,
\ben
\hat{\theta}_{n,\Gamma}
&=&
    \Gamma_n(G)^{-1}
    [\mu_n(F(G+PG))-
    \mu_n(F)\mu_n(G+PG)]\\
\;\;\;\;
\hat{\theta}_{n,\K}
&=&
    \K_n(G)^{-1}
    [\mu_n(F(G+PG))-
    \mu_n(F)\mu_n(G+PG)],
\een
where the empirical $k\times k$
matrices $\Gamma_n(G)$ and $\K_n(G)$ are
defined, respectively, by,
\ben
(\Gamma_n(G))_{ij}
&=&
    \mu_n(G_iG_j)-\mu_n((PG_i)(PG_j))\\
\mbox{and}
\;\;\;\;
(\K_n(G))_{ij}
&=&
\frac{1}{n-1}\sum_{t=1}^{n-1}
(G_i(X_t)-PG_i(X_{t-1})) (G_j(X_t)-PG_j(X_{t-1})).
\een

\medskip

\noindent
{\bf Theorem 3. }
{\em
Suppose that the chain $\{X_n\}$ is reversible and
satisfies conditions {\em (A)}. If the functions
$F,G$ are both in $L_\infty^W$
and the matrix $\Gamma(G)$ is nonsingular,
then both of the estimators for
$\theta^*$ are a.s.-consistent:
\ben
\hat{\theta}_{n,\Gamma}\to\theta^*
&&
    \;\;\;\mbox{a.s., as }\;n\to\infty;\\
\hat{\theta}_{n,\K}\to\theta^*
&&
    \;\;\;\mbox{a.s., as }\;n\to\infty.
\een
}

\medskip

Recall the definitions of the two estimators
$\mu_{n,\Gamma}(F)$ and $\mu_{n,\K}(F)$
in equations~(\ref{eq:multi-est2-2})
and~(\ref{eq:multi-est2}), 
from Sections~\ref{s:thetastar} and~\ref{s:Gamma}, 
respectively.
Combining the two theorems, yields the desired
asymptotic properties of the two estimators:

\medskip

\noindent
{\bf Corollary 1. }
{\em Suppose that the chain $\{X_n\}$ is reversible
and satisfies conditions {\em (A)}. If the functions
$F,G$ are both in $L_\infty^W$
and the matrix $\Gamma(G)$ is nonsingular,
then the modified estimators
$\mu_{n,\Gamma}(F)$
and $\mu_{n,\K}(F)$
for $\pi(F)$ satisfy:
\begin{enumerate}
\item[{\bf (i)}]
{\sc [LLN] }
The modified estimators
$\mu_{n,\Gamma}(F)$,
$\mu_{n,\K}(F)$
both converge to $\pi(F)$ a.s.,
as $n\to\infty$.
\item[{\bf (ii)}]
{\sc [CLT] } If $\sigma_{F_{\theta^*}}^2:=
\pi(\hat{F}_{\theta^*}^2-(P\hat{F}_{\theta^*})^2)$ is nonzero, then
the normalized modified averages
$\sqrt{n}[\mu_{n,\Gamma}(F)-\pi(F)]$ and
$\sqrt{n}[\mu_{n,\K}(F)-\pi(F)]$ converge in distribution to
$N(0,\sigma_{F_{\theta^*}}^2)$, as $n\to\infty$, where the variance
$\sigma_{\theta^*}^2$ is minimal among all estimators based on the
control variate $U=G-PG$, in that
$\sigma_{\theta^*}^2=\min_{\theta\in\RL^k} \sigma_{\theta}^2.$
\end{enumerate}
}

\medskip

Finally we turn to the batch-means estimators 
of Section~\ref{s:batch}. Recall 
the definitions of the estimators 
$\tilde{\theta}_{n,M}$
and $\tilde{\mu}_{n,M}(F)$ in equations~(\ref{eq:thetabatch}) 
and~(\ref{eq:mubatch}), respectively.
Our next result shows that 
$\tilde{\theta}_{n,M}$ converges 
to $\tilde{\theta}_M$ defined in~(\ref{eq:thetaM}),
and gives an a.s.-law of large numbers result
and a corresponding central limit
theorem for the estimators 
$\tilde{\mu}_{n,M}(F)$.
Its proof follows along the same line
as the proofs of the corresponding statements
in Theorem~3 and Corollary~1.
[Note that in the one-dimensional 
setting of Corollary~2 the assumption
that $\Gamma(G)$ is nonsingular reduces
to assuming that $\sigma_U^2=\pi(G^2-(PG)^2)$
is nonzero.]

\medskip

\noindent
{\bf Corollary 2. } {\em Under the assumptions
of Theorem~3, for any fixed $M\geq 0$, 
as $n\to\infty$ we have:
\begin{enumerate}
\item[{\bf (i)}] {\em [LLN]}
$\;\tilde{\theta}_{n,M}\to \tilde{\theta}_M$ 
a.s., and $\tilde{\mu}_{n,M}(F)\to \pi(F)$ a.s.
\item[{\bf (ii)}] {\em [CLT]}
$\;\sqrt{n}[\tilde{\mu}_{n,M}(F)-\pi(F)]\tod N(0,\tau_M^2)$,
where the variance $\tau_M^2$ is given 
by,
\be
\tau_M^2=
\sigma_F^2+\tilde{\theta}_M^2\sigma_U^2
-2\tilde{\theta}_M\sum_{n=-\infty}^\infty\COV_\pi(F(X_0),U(X_n)).
\label{eq:tauSM}
\ee
\end{enumerate}
}

Some additional results on the long-term behavior of estimators
similar to the ones considered above can be found in
\citet{meyn:06}, \citet[Chapter~11]{meyn:book},
and finer asymptotics (including large deviations
bounds and Edgeworth expansions) 
can be derived under stronger assumptions
from the results in 
\cite{kontoyiannis-meyn:Itmp}, 
\cite{kontoyiannis-meyn:II}.


\section{Concluding Remarks and Further Extensions}
\label{s:conclude}

{\em Summary of results. }
This work introduced a general methodology
for the construction and effective application 
of control variates to estimation problems
with MCMC data. The starting point was the
observation by \citet{henderson:phd} that,
for an arbitrary function $G$ on the state 
space of a chain with transition kernel $P$,
the function $U:=G-PG$ has zero mean with 
respect to the stationary distribution $\pi$
and can thus be used as a control 
variate. The two main issues treated here
are:
($i$)~The selection of {\em basis functions} $G=\{G_j\}$ 
for the construction of effective control variates 
$U_j=G_j-PG_j$; and ($ii$)~The problem of effectively
and consistently estimating the 
optimal coefficients $\theta^*=\{\theta_j^*\}$ 
of the linear combination $\sum_j\theta_j^*U_j$.
The main difficulty
was identified in Sections~\ref{s:cv}
and~\ref{s:thetastar} as stemming from the fact that 
the obvious answer to both of these issues
involves the solution $\hat{F}$ of an associated 
Poisson equation. Since $\hat{F}$ is well-known 
to be notoriously difficult to compute
and only known explicitly in a handful
of very simple examples, 
alternative approaches were necessarily sought.

For reversible chains, in Section~\ref{s:thetastar} 
we derived new representations of $\theta^*$ 
that do not involve $\hat{F}$, and in Section~\ref{s:basic}
we derived the exact solution of the Poisson equation
for a specific MCMC scenario. These theoretical
results motivated the {\em basic methodology}
for variance reduction proposed in 
Section~\ref{s:basic}. 
In Section~\ref{s:simple} it was
applied to three representative MCMC examples,
demonstrating that 
the resulting reduction in the estimation variance 
is generally quite significant.
Extensions in several 
directions were developed in Section~\ref{s:generalMeth}, 
in each case 
illustrated via a simulation 
experiment of Bayesian inference via MCMC.

In Section~\ref{s:issues}, the most common
estimator for $\theta^*$ that has been 
used in the literature -- based on the
method of batch-means -- was examined, and the
resulting estimator for $\pi(F)$ was shown
to be both computationally expensive and 
generally rather ineffective, often severely so. 
This was demonstrated analytically in Proposition~3
and also empirically via MCMC examples. 
Section~\ref{s:issues} also contains a brief discussion 
of two alternative approaches for estimating $\theta^*$
consistently. Finally, all methodological and asymptotic 
arguments were rigorously justified in Section~\ref{s:theory},
under easily verifiable and essentially minimal 
conditions. 

\medskip

\noindent
{\em Applicability. } 
One of the strengths of our approach to the use of
control variates in MCMC estimation is that, unlike 
in the classical case of independent 
sampling where control variates need to be identified 
in an {\em ad hoc} fashion for each specific application,
the basic methodology developed here (and its extensions)
is immediately applicable to a wide range of MCMC 
estimation problems. The most natural class of such
problems consists of all Bayesian inference studies where 
samples from the posterior are generated 
by a conjugate random-scan 
Gibbs sampler. Recall that conjugate Gibbs sampling 
is the key ingredient in,
among others: Bayesian inference for dynamic 
linear models, e.g., \citet{reis}; 
applications of slice Gibbs with auxiliary variables, e.g., \citet{damien99}; 
Dirichlet processes, e.g., \citet{MacEachernmuller};
and spatial regression models, e.g., \citet{gamerman03}.

More generally, the present methodology applies to 
any MCMC setting satisfying the following two requirements:
That the chain be reversible and that the
conditional expectations $PG(x)=E[G(X_{n+1})|X_n=x]$ 
are explicitly computable for some simple functions $G$. 
There is a large collection of samplers with these 
properties, including certain versions of hybrid 
Metropolis-within-Gibbs algorithms (as in Example~3), 
certain Metropolis-Hastings samplers on discrete states 
spaces (as in Example~6), and Markovian models 
of stochastic networks (as in \citet{meyn:book}).
To ensure that these two requirements are
satisfied, most of the experiments reported in 
Sections~\ref{s:simple},~\ref{s:generalMeth}
and~\ref{s:issues} were performed using the
{\em random-scan} version of Gibbs sampling. 
This choice is not {\em a priori}
restrictive since the convergence properties 
of random-scan algorithms are generally comparable 
(and sometimes superior) to those of systematic-scan 
samplers; see, e.g., the discussions in 
\citet{diaconis-ram}, \citet{robsah97}.  

We also observe that, as the present methodology
is easily implemented as a post-processing 
algorithm and does not interfere in the actual
sampling process, any implementation technique that 
facilitates or accelerates the MCMC convergence 
(such as blocking schemes, transformations, 
other reversible chains, and so on), can be 
used, as long as reversibility is maintained.

\medskip

\noindent
{\em Further extensions. } Perhaps the most 
interesting class of MCMC samplers that could be 
considered next is that of general Metropolis-Hastings 
algorithms.  When the target distribution is discrete or, 
more generally, when the proposal distribution is 
discrete and the number of possible moves is not 
prohibitively large, then our control variates
methodology can be used as illustrated in Example~6. 
But in the case of general, typically continuous 
or multidimensional proposals, there is a basic obstacle:
The presence of the accept/reject probability 
in each step makes it impossible to compute
the required 
conditional expectation $PG(x)$
in closed form, for any $G$.
If we consider the extended chain $\{(X_n,Y_n)\}$
that includes the values of the proposed moves $Y_n$ 
(as done, e.g., by \citet{Hammer08}
and \citet{delmas-jourdain}),
then the computation of $PG$ is straightforward
for any function $G(x,y)$ that only depends on $x$;
but the chain $\{(X_n,Y_n)\}$ is no longer 
reversible, and there are no clear
candidates for good basis functions $G$.
A possibly more promising point of view
is to consider the computation of $PG$
an issue of numerical integration, and to 
try to estimate the required values $PG(X_n)$
based on importance sampling or any one 
of the numerous standard numerical integration 
techniques; these considerations are well
beyond the scope of the present work.

In a different direction, an interesting and
potentially useful point would be to examine
the effect of the use of control variates in
the estimation {\em bias}.
Although the variance of the standard
ergodic averages $\mu_n(F)$ is a ``steady-state''
object, in that it characterizes their long-term behavior 
and depends neither on the initial condition 
$X_0 = x$ nor on the transient behavior of the chain,
the bias depends heavily on the initial condition
and it vanishes asymptotically. Preliminary
computations as in the 
unpublished notes of \citet{dellaportas-K:arxiv}
indicate that the bias of $\mu_n(F)$
decays to zero approximately like $\hat{F}(x)/n$, 
and that,
using on a single control variate $U = G- PG$
based on a function $G\approx\hat{F}$, can significantly
reduce the bias. It would be interesting in future
work to compute the coefficient vector $\theta^b$ 
which minimizes the bias of $\mu_n(F_\theta)$
for a given collection of basis functions $\{G_j\}$, 
to study ways in which $\theta^b$ can be estimated
empirically,
and to examine the effects that the use of 
$\theta^b$ in conjunction with the control
variates $U=G-PG$ would have on the 
variance of the resulting estimator
for $\pi(F)$.

A final point which may merit further attention is the
potential problem of including too many control 
variates in the modified estimator $\mu_{n,\K}(F)$.
This issue has been studied extensively in the classical
context of estimation based on i.i.d.\ samples;
see for example,
\citet{lavenberg-welch},
\citet{law-kelton:book},
\citet{nelson:04},
\citet[pp.~200-202]{glasserman:book},
\citet{caris:05}.
Since the optimal coefficient vector $\theta^*$ is not known a priori, 
using many control variates may in fact increase the variance of the 
modified estimators $\mu_{n,\K}(F)$ relative to $\mu_n(F)$,
and care must be taken to ensure that the most effective 
subset of all available control variates is chosen. 
Common sense suggests that the values of all the estimated 
parameters in the vector $\hat{\theta}_{n,\K}$ should 
be examined, and the control variates corresponding 
to coefficients that are approximately zero should 
be discarded. And since the MCMC output consists 
of simulated data from a known distribution, it may 
be possible to do this in a systematic fashion by 
developing a classical hypothesis testing procedure.


\section*{Acknowledgments}
We are grateful to Sean Meyn and Zoi Tsourti for many interesting
conversations related to this work, and to Persi Diaconis
and Christian Robert for insightful comments on an earlier
version of this paper.

\appendix
\section*{Appendix: Proofs of Theorems~2,~3 and Corollaries~1,~2}

\noindent
{\sc Proof of Theorem 2. } Since any small set is
petite, \citet[Section~5.5.2]{meyn-tweedie:book2},
the $f$-norm ergodic theorem of
\citet{meyn-tweedie:book2} implies that
$\{X_n\}$ is positive recurrent with
a unique invariant measure $\pi$
such that (\ref{eq:V3bound}) holds,
and \citet[Theorem~11.3.4]{meyn-tweedie:book2}
proves the Harris property, giving~(i).

From \citet[Theorem~14.0.1]{meyn-tweedie:book2}
we have that, under (V3), $\pi(W)<\infty$.
Since $F$ is in $L_\infty^W$, $\pi(|F|)$ is finite,
and since $G\in L_\infty^W,$ Jensen's inequality
guarantees that $\pi(|U|)$ is finite. The invariance
of $\pi$ then implies that $\pi(U)=0$;
therefore,
\citet[Theorem~17.0.1]{meyn-tweedie:book2}
shows that $\mu_n(F)\to\pi(F)$ and $\mu_n(U)\to 0$
a.s.\ as $n\to\infty$, and since
$\theta_n\to\theta$ by assumption,
$\mu_n(F_\theta)$ also converges
to $\pi(F)$ a.s., proving~(ii).

The existence of a solution $\hat{F}$ to
the Poisson equation in~(iii) follows from
\citet[Theorem~17.4.2]{meyn-tweedie:book2},
and its uniqueness from
\citet[Theorem~17.4.1]{meyn-tweedie:book2}.
The CLT in~(iv) is a consequence of
\citet[Theorem~17.4.4]{meyn-tweedie:book2}.

Finally, since $F,G\in L_\infty^W$,
the functions $U$ and $F_\theta$ are in
$L_\infty^W$ too, so $\hat{U}_j$ and
$\hat{F}_\theta$ exist for each $j=1,2,\ldots,k$.
As in~(iv), the scaled averages
$\sqrt{n}[\mu_n(F_\theta)-\pi(F)]$
and $\sqrt{n}\mu_n(U_j)$
converge in distribution to
$N(0,\sigma_{F_\theta}^2)$ and
$N(0,\sigma_{U_j}^2)$,
respectively, for each $j$,
where the variances
$\sigma_{F_\theta}^2$
and $\sigma_{U_j}^2$
are as in~(iii).
Writing $\theta=(\theta_1,\theta_2,\ldots,\theta_k)^t$
and
$\theta_n=(\theta_{n,1},\theta_{n,2},\ldots,\theta_{n,k})^t$,
we can express,
$$\sqrt{n}[\mu_n(F_{\theta_n})-\pi(F)]
= \sqrt{n}[\mu_n(F_\theta)-\pi(F)]+
\sum_{j=1}^k\Big\{
(\theta_{n,j}-\theta_j)
\sqrt{n}\mu_n(U_j)
\Big\}.$$
Each of the terms in the
second sum on the right-hand-side
above converges to zero in probability,
since and $\sqrt{n}\mu_n(U_j)$ converges
to a normal distribution and
$(\theta_{n,j}-\theta_j)\to 0$ a.s.
Therefore, the sum converges to zero
in probability, and the CLT in~(v)
follows from (iv).
\qed


Note that the assumption $\sigma_{U_j}^2\neq 0$ in
the theorem is not necessary, since the case
$\sigma_{U_j}^2=0$ is trivial in view
of \citet[Proposition~2.4]{kontoyiannis-meyn:Itmp},
which implies that, then,
$\sqrt{n}\mu_n(U_j)\to 0$ in probability,
as $n\to\infty$.

\medskip

\noindent
{\sc Proof of Theorem 3. }
We begin by justifying the computations in Sections~\ref{s:cv2}
and~\ref{s:thetastar}.
Define
$\sigma_{F_\theta}^2=\pi(\hat{F}_\theta^2- (P\hat{F}_\theta)^2)$,
where $\hat{F}$ exists by Theorem~2.
Since $\hat{F}$ solves the Poisson equation for $F$,
it is easy to check that $\hat{F}_\theta:=\hat{F}-\langle\theta,G\rangle$
solves the Poisson equation for $F_\theta$.
Substituting this in the above expression
for $\sigma_{F_\theta}^2$ yields (\ref{eq:varTheta}).
To see that all the functions in (\ref{eq:varTheta})
are indeed integrable recall that
$\hat{F}\in L_\infty^{V+1}$
and note that,
since $V$ is nonnegative, (V3) implies that
$1\leq W\leq V + b\IND_C$, hence
$\pi(W^2)$ is finite since $\pi(V^2)$
is finite by assumption.
Therefore, since $G\in L_\infty^{W}$,
$\hat{F}$ and $G$ are both in $L_2(\pi)$,
and H\"{o}lder's inequality implies that
$\pi(\hat{F}\langle\theta,G\rangle)$
is finite. Finally, Jensen's inequality
implies that $P\hat{F}$ and $PG$ are also
in $L_2(\pi)$, so that $\pi(P\hat{F}\langle\theta,PG\rangle)<\infty$.
And, for the same reasons, all the functions
appearing in the computations leading to the results
of Propositions~1 and~2 are also integrable.

The expression for the
optimal $\theta^*$ in (\ref{eq:thetaStar-k})
is simply the solution for the minimum of the
quadratic in (\ref{eq:varTheta}). Again,
note that $\hat{F},G,P\hat{F}$ and $PG$
are all in $L_2(\pi)$ so $\theta^*$
is well-defined.

The consistency proofs follow from repeated
applications of the ergodic theorems established
in Theorem~2. First note that, since
$G\in L_\infty^W$ and $\pi(W^2)<\infty$ as
remarked above, the product $G_iG_j$ is
$\pi$-integrable, and by Jensen's inequality
so is any product
of the form $(PG_i)(PG_j)$. Therefore,
the ergodic theorem of \citet[Theorem~17.0.1]{meyn-tweedie:book2}
implies that $\Gamma_n(G)\to\Gamma(G)$ a.s.
Similarly, the functions $F$, $G$, $PG$, $FG$ and $FPG$
are all $\pi$-integrable, so that the same ergodic
theorem implies that $\hat{\theta}_{n,\Gamma}$
indeed converges to $\theta^*$ a.s., as $n\to\infty$.

To establish the corresponding result for
$\hat{\theta}_{n,\K}$, it suffices to show that
$K_n(G)\to K(G)$ a.s., and to that end we consider
the bivariate chain $Y_n=(X_n,X_{n+1})$ on the
state space $\state\times\state$. Since
$\{X_n\}$ is $\psi$-irreducible and aperiodic,
$\{Y_n\}$ is $\psi^{(2)}$-irreducible and aperiodic
with respect to the bivariate measure
$\psi^{(2)}(dx,dx'):=\psi(dx)P(x,dy)$.
Given functions $W,V$ a small set $C$ and
a constant $b$ so that (V3) holds, it is
immediate that (V3) also holds for $\{Y_n\}$
with respect to the functions
$V^{(2)}(x,x')=V(x')$,
$W^{(2)}(x,x')=W(x')$, the small
set $\state\times C$, and the same $b$.
The unique invariant measure of
$\{Y_n\}$ is then
$\pi^{(2)}(dx,dx'):=\pi(dx)P(x,dy)$,
and $\pi^{(2)}((V^{(2)})^2)$ is finite.
Therefore, assumptions (A) hold for
$\{Y_n\}$ and, for each pair
$1\leq i,j\leq k$ we can invoke the
ergodic theorem
\citet[Theorem~17.0.1]{meyn-tweedie:book2}
for the $\pi^{(2)}$-integrable function,
$$H(x,x'):=
(G_i(x')-PG_i(x)) (G_j(x')-PG_j(x)),
$$
to obtain that, indeed, $\K_n(G)\to \K(G)$ a.s.
\qed


\noindent
{\sc Proof of Corollary~1. }
The ergodic theorems in~(i) are immediate
consequences of Theorem~2~(ii) combined
with Theorem~3. The computation
in Section~\ref{s:cv2} which shows that
$\theta^*$ in (\ref{eq:thetaStar-k})
indeed minimizes $\sigma_{F_\theta}^2$
(justified in the proof of Theorem~3)
shows that $\sigma_{\theta^*}^2=\min_{\theta\in\RL^k}
\sigma_{\theta}^2.$
Finally, the assumption that $\Gamma(G)$
is nonsingular combined with Proposition~1,
imply that
all the variances $\sigma_{U_j}^2$ must be
nonzero. Therefore, Theorem~3 combined
with the central limit theorems
in parts~(iv) and~(v) of Theorem~2, prove
part~(ii) of the Corollary.
\qed

\noindent
{\sc Proof of Corollary 2. }
The a.s.-convergence statements in~(i) 
follow the ergodic theorem,
as in the proofs of Theorems~2 and~3.
The a.s.-convergence of the 
denominator of (\ref{eq:thetabatch}),
$\mu_n(G^2-(PG)^2)\to
\pi(G^2-(PG)^2)$, is a special case
(corresponding to $k=1$) of the
a.s.-convergence of $\Gamma_n(G)$
to $\Gamma(G)$ proved in Theorem~3.
Considering the $(2M+1)$-variate chain
instead of the bivariate chain as in the
proof of Theorem~3, we can apply 
the ergodic theorem with the same
integrability assumptions,
to obtain that the sum in the numerator
of (\ref{eq:thetabatch}) converges
a.s.\ to $\sum_{|j|\leq M} E_\pi[F(X_0)U(X_j)]$,
proving that $\tilde{\theta}_{n,M}\to\tilde{\theta}_M$
a.s., as $n\to\infty$.
For the modified averages, note that 
$\tilde{\mu}_{n,M}(F)$ is simply
$\mu_n(F_{\tilde{\theta}_{n,M}})$.
Then the LLN and CLT results for
$\tilde{\mu}_{n,M}(F)$ follow from
parts~(ii) and~(v) of Theorem~2, respectively.
Finally, the limiting variance
$\tau_M^2$ equals $\sigma^2_{\tilde{\theta}_M}$,
which, using the representation in 
equation~(\ref{eq:varseries}),
can be expressed as claimed in~(\ref{eq:tauSM}).
\qed

\renewcommand{\baselinestretch}{0} 

{\small
\def\cprime{$'$}

}

\end{document}